\documentclass[%
 reprint,
superscriptaddress,
 amsmath,amssymb,
 aip,
floatfix,
]{revtex4-1}

\usepackage{import}

\usepackage{graphicx}
\usepackage{dcolumn}
\usepackage{bm}
\usepackage{siunitx}
\usepackage[scr=euler]{mathalfa}
\usepackage{enumitem}

\DeclareFontFamily{U}{BOONDOX-calo}{\skewchar\font=45 }
\DeclareFontShape{U}{BOONDOX-calo}{m}{n}{
  <-> s*[1.05] BOONDOX-r-calo}{}
\DeclareFontShape{U}{BOONDOX-calo}{b}{n}{
  <-> s*[1.05] BOONDOX-b-calo}{}
\DeclareMathAlphabet{\mathproper}{U}{BOONDOX-calo}{m}{n}
\SetMathAlphabet{\mathproper}{bold}{U}{BOONDOX-calo}{b}{n}
\DeclareMathAlphabet{\mathbproper}{U}{BOONDOX-calo}{b}{n}
\DeclareMathAlphabet{\mathpzc}{OT1}{pzc}{m}{it}
\newcommand\christoffel[2]{\genfrac{ \{ }{ \}  }{0pt}{1}{#1}{#2} }
\usepackage{xcolor}

\begin{document}

\preprint{APS/123-QED}

\title{Static energetics in gravity}

\author{W E V Barker}
 \email{wb263@mrao.cam.ac.uk}
\affiliation{%
  Astrophysics Group, Cavendish Laboratory, JJ Thomson Avenue, Cambridge CB3 0HE, UK
}%
\affiliation{Kavli Institute for Cosmology, Madingley Road, Cambridge CB3 0HA, UK}

 \author{A N Lasenby}%
\email{a.n.lasenby@mrao.cam.ac.uk}
\affiliation{%
  Astrophysics Group, Cavendish Laboratory, JJ Thomson Avenue, Cambridge CB3 0HE, UK
}%
\affiliation{Kavli Institute for Cosmology, Madingley Road, Cambridge CB3 0HA, UK}

\author{M P Hobson}%
\email{mph@mrao.cam.ac.uk}
\affiliation{%
  Astrophysics Group, Cavendish Laboratory, JJ Thomson Avenue, Cambridge CB3 0HE, UK
}%
\author{W J Handley}%
\email{wh260@mrao.cam.ac.uk}
\affiliation{%
  Astrophysics Group, Cavendish Laboratory, JJ Thomson Avenue, Cambridge CB3 0HE, UK
}%
\affiliation{Kavli Institute for Cosmology, Madingley Road, Cambridge CB3 0HA, UK}

\date{\today}

\begin{abstract}

  A stress-energy tensor, $\tau_{ab}$, for linear gravity in the physical spacetime, $\mathcal{M}$, approximated by a flat background, $\check{\mathcal{M}}$, and adapted to the harmonic gauge, was recently proposed by Butcher, Hobson and Lasenby. By removing gauge constraints and imposing full metrical general relativity, we find a natural generalisation of $\tau_{ab}$ to the pseudotensor of Einstein, $_Et_{ab}$. M{\o}ller's pseudotensor, $_Mt_{ab}$, is an alternative to $_Et_{ab}$ formulated using tetrads, and is thus naturally adapted to e.g. Einstein-Cartan gravity. Gauge theory gravity uses geometric algebra to reproduce Einstein-Cartan gravity, and is a Poincar\'{e} gauge theory for the spacetime algebra: the tetrad and spin connection appear as gauge fields on Minkowski space, $M_4$. We obtain the pseudotensor of M{\o}ller for gauge theory gravity, $_M\mathsf{t}(a)$, using a variational approach, also identifying a potentially interesting recipe for constructing conserved currents in that theory.
  We show that in static, spherical spacetimes containing a gravitational mass $M_T$ the pseudotensors in the spacetime algebra, $_M\mathsf{t}(a)$ and $_E\mathsf{t}(a)$, describe gravitational stress-energy as if the gravitational potential were a scalar (i.e. Klein-Gordon) field, $\varphi$, coupled to gravitational mass density, $\varrho$, on the Minkowski background $M_4$. The old Newtonian formula $\varphi=-M_T/r$ successfully describes even strong fields in this picture. The Newtonian limit of this effect was previously observed in $\tau_{ab}$ on $\check{\mathcal{M}}$ for linear gravity. We also draw fresh attention to the conserved mass of a static system, $\mathscr{M}_T<M_T$. When compared with either $\varrho$ or the density of the Komar mass in the Newtonian limit, $\mathscr{M}_T$ produces a local virial theorem -- such behaviour is usually associated with the proper mass of the system, $\mathproper{M}_T>M_T$. We observe that the gravitational energy of Einstein and M{\o}ller add to $\mathscr{M}_T$ on $M_4$ to give $M_T$. We demonstrate the Klein-Gordon correspondence and mass functions using the `Schwarzschild star' solution for an incompressible perfect fluid ball.
\end{abstract}

\pacs{04.20.Cv, 04.50.Kd}
\maketitle

\section{Introduction}\label{s1}
It is widely accepted that the energy contained in a gravitational field cannot in general be localised. 
This paradigm, which developed over the century following the advent of general relativity\footnote{For an comprehensive discussion of the `energy problem' in general relativity, see for example the review by J. Goldberg\cite{jiri_1}.}, is often regarded as a consequence of the equivalence principle. 
It is equally well accepted that gravitational fields carry such an energy in the first place. The ebb and flow of energy-momentum between matter and gravity explains the emission and recent detection of gravitational waves, with both processes mediated by the covariant conservation of the stress-energy tensor of matter
\begin{eqnarray}\label{origin}
  \nabla_a T_b^a=0.
\end{eqnarray}
Counter-intuitively, this law states that the energy and momentum of matter, whose densities are the contractions of $T^a_b$ with an observer's four-velocity, are \textit{not} generally conserved in the presence of a gravitational field. Many introductory texts (e.g. \onlinecite{wald, landau}) are quick to observe such energy-momentum exchange is not inevitable through the generic counterexample: a stationary spacetime is furnished with a global timelike Killing vector, $K^a$, so the quantity
\begin{eqnarray}\label{consone}
  Q_T=\int_{\Sigma_t}\epsilon_{abcd}T^{ae}K_e
\end{eqnarray}
is independent of the Cauchy hypersurface $\Sigma_t$ used to define it, and hence constitutes a conserved charge.
\par
In contrast to energy localisation, the global picture of gravitational energetics is often less ambiguous. This is the case for many spacetimes of astrophysical interest, which can be `patched on' to the universe at large because they are asymptotically flat. In such cases the Newtonian regime at spatial infinity provides an observer with a clear account of the total gravitational mass, $M_T$, of the system. 
Komar \cite{komar} proposed a derived quantity for stationary asymptotically flat systems, designed to agree with precisely this Newtonian value. 
The integral theorem can be used to obtain a Komar mass `density' which is proportional to the Ricci tensor. Such a picture, in which $M_T$ is exclusively distributed wherever $T^{ab}$ is nonvanishing, is dissatisfying because it does not reflect the common assumption that part of the gravitational mass is locked away in the gravitational field, a view explored more recently by Katz, Lynden-Bell and Bi\v{c}\'{a}k\cite{jiri_2,jiri_3}.
The notion of asymptotic flatness was developed during the golden age of general relativity to imply conformal isometry to a bounded region of some curved, nonphysical spacetime (in the case of Minkowski spacetime, which clearly has asymptotic flatness, the nonphysical spacetime is Einstein's static universe).
This enabled the mass of Komar to be generalised to that of Bondi \cite{bondi} which is evaluated at various sections of null infinity\footnote{Bondi's mass will not be of use to us because we do not consider radiating systems.}.
At spatial infinity, the Hamiltonian formulation of general relativity attributed to Arnowitt, Deser and Misner \cite{adm} provides a further definition of $M_T$ reliant on asymptotically Cartesian coordinates.
\par
For those dissatisfied with the global picture, attempts to localise gravitational energetics usually take the form of energy-momentum \textit{complexes} -- objects of questionable gauge invariance which emulate a combined stress-energy tensor for matter and gravity. A basic requirement of a complex following from our discussion is that it integrates in some sense to give $M_T$. 
A further requirement is that it be identically conserved, as with the matter stress energy tensor in special relativity. Conservation is built in by defining the complex to be the gradient of a superpotential constructed from the dynamical variables of gravity. 
The first complex  $_E\theta_a^{\ b}$ was proposed by Einstein \cite{einstein} in 1916, though the corresponding superpotential is attributed to Freud \cite{freud}, 
\begin{eqnarray}
  _E\theta_a^{\ b}={\partial_c} {_F\Psi_{a}^{\ bc}},
\end{eqnarray}
where Freud's superpotential $_F\Psi_a^{\ bc}$ is a function of the metric and its first derivatives and is skew-symmetric in its final pair of indices. The utility of the `special' conservation law
\begin{eqnarray}
  {\partial_b} {_E\theta_{a}^{\ b}}=0,
\end{eqnarray}
is evident when the field equations are used to collect the second derivatives of the metric appearing in $_E\theta_a^{\ b}$ into $T_a^b$, so partitioning the energetics of matter and gravity
\begin{eqnarray}\label{einstein_comp}
  _E\theta_a^{\ b}=\sqrt{-g}(T_a^b+{_Et_a^{\ b}}).
\end{eqnarray}
The remaining quantity $_Et_a^{\ b}$ is known as \textit{Einstein's pseudotensor}. Over the four decades following the introduction of \eqref{einstein_comp}, the evident freedom in the choice of superpotential led many authors to develop their own complexes, including Landau and Lifshitz \cite{landau}, Komar and M{\o}ller. During this time, the scope of the geometric theory of gravitation was expanded from the general relativity of Riemann space, $V_4$, to the Einstein-Cartan theory of Riemann-Cartan space, $U_4$. This resulted in the title of `dynamical variable of gravity' passing from the metric to its `square root', which is the tetrad or vierbein.
Having encountered difficulties with his early metrical attempts, M{\o}ller \cite{moller1} constructed a new superpotential ${_M\Psi_a^{\ bc}}$ from the tetrad and its first derivatives. 
Unlike its predecessors, M{\o}ller's superpotential is a tensor under the usual diffeomorphisms of the spacetime, but not under Lorentz rotations of the tetrads. The corresponding energy-momentum complex
\begin{eqnarray}\label{moller_comp}
  _M\theta_a^{\ b}=\sqrt{-g}(T_a^b+{_Mt_a^{\ b}}),
\end{eqnarray}
and pseudotensor are otherwise fairly analogous to those of Einstein.
\par
In the context of our opening remarks, it is not surprising that energy-momentum complexes suffer greatly under the principle of equivalence. Many authors have objected that upon falling, the associated pseudotensors promptly vanish along with their local account of gravitational energy. Consequently their deployment is usually confined to privileged quasi-Cartesian coordinate systems, though this is at least compatible with the techniques of the Hamiltonian formulation at spatial infinity.
\par
Just as the Newtonian regime provides a valuable global concept of gravitational energy, so it has proven useful in energy localisation: from the perspective of linear gravity on a flat background, pseudotensors and tensors are indistinguishable.
Bi\v{c}\'{a}k and Schmidt \cite{Bicak} have charted the freedom and ambiguity that is to be found at lowest perturbative order in the construction of gravitational stress-energy tensors. 
Their analysis includes a symmetric tensor $\tau_{ab}$ developed in a recent paper by Butcher, MPH and ANL \cite{butcher2}. 
If the physical spacetime, $\mathcal{M}$, is an example of $V_4$, this tensor is constructed in the background spacetime $\check{\mathcal{M}}$ (which is nothing more than Minkowski space, $M_4$), to account for the local non-conservation of matter energy-momentum implied by \eqref{origin}.
In the harmonic gauge, it is the unique symmetric tensor to do so. A curious observation made in \onlinecite{butcher2} is that $\tau_{ab}$ treats the Newtonian gravitational potential as if it were a matter-generated Klein-Gordon field, but as with any linear effect this correspondence should not be over-interpreted.
In \onlinecite{butcher3} a similar procedure led to a tensor for gravitational spin. 
The same authors demonstrated in \onlinecite{butcher4} that these tensors are the canonical Noether currents in Einstein-Cartan gravity under a perturbative expansion of the Einstein-Cartan Lagrangian which they developed in \onlinecite{butcher1}. 
\par
The succession of Einstein-Cartan theory was (and largely remains) formal, with the vast majority of the literature addressing general relativity. 
Nevertheless, the advent of the tetrad and spin connection eventually gave rise to a rich new class of \textit{gauge theories} of gravity.
The Poincar\'{e} group was fully gauged by Kibble\footnote{Kibble's work\cite{kibble} follows early efforts by Utiyama and concurrently with Sciama -- see Section I of \onlinecite{lasenby4} and the references therein.} who considered an action analogous to that of Einstein and Hilbert in which the gravitational gauge fields are minimally coupled to matter. The unifying mathematical language of geometric algebra has also been brought to bear on the problem by gauging in the geometric algebra of Minkowski spacetime, or \textit{spacetime algebra}. During the procedure, there arise natural ways to implement minimal coupling and an Einstein-Hilbert action. The result \cite{lasenby, doran}, known as \textit{gauge theory gravity}, appears quite alien when compared to Kibble's theory, but both may be re-interpreted geometrically\footnote{Modulo topological effects.} as Einstein-Cartan theory in which torsion is sourced by material spin.
\par
There are two common alternative approaches to constructing energy-momentum complexes. Rather than composing the relevant superpotential from the beginning, it may prove efficient to isolate it by `splitting' the Einstein equations. 
Hestenes \cite{hestenes} has demonstrated that this method lends itself very strongly to gauge theory gravity in the spacetime algebra, where the Einstein tensor can be written in his \textit{unitary form}. 
Accordingly he obtains the complexes, or splits, of Einstein, Landau-Lifshitz and M{\o}ller, along with one which is original. 
The other method is that referred to by M{\o}ller as \textit{variational}: it may be possible to construct an alternative Lagrangian to that of Einstein and Hilbert, from which the required energy-momentum complex follows as an (affine) canonical stress-energy tensor. M{\o}ller employed both variational and superpotential methods when proposing \eqref{moller_comp}, whilst the variational approach was suggested by Einstein for \eqref{einstein_comp}. 
Unlike the geometric theories of M{\o}ller's day, gauge theory gravity was developed from the very beginning as a Lagrangian field theory, so one would expect it to be well suited to the variational method.
\par
\vspace{2mm}
We use the methods of gauge theory gravity and the spacetime algebra to provide a fresh perspective on the localisation of $M_T$ and the role of the conserved charge $Q_T$ -- we confine our discussion to static spacetimes containing perfect fluids without spin. In the ensuing absence of torsion, gauge theory gravity may be geometrically interpreted as general relativity. In our treatment, we will relate the formalisms of Butcher, Einstein and M{\o}ller.
\par
The remainder of this article is set out as follows. Section \ref{s2} pertains to general relativity. In Section \ref{s2a} through to \ref{s2d} we review the approach of Butcher et al and see how it might be extended to nonlinear gravity. In Section \ref{s2e} we make some observations on the relativistic mass of static, spherically symmetric perfect fluids. 
\par
Section \ref{s3} addresses some of the issues raised in Section \ref{s2} using the gauge theory approach, beginning with a brief introduction to gauge theory gravity. In Sections \ref{s3a} and \ref{s3b} we discuss energy localisation formalisms in gauge theory gravity as obtained through the variational approach, in particular the pseudotensor of M{\o}ller. This enables us to generalise the Klein-Gordon correspondence of $\tau_{ab}$ in Section \ref{s3c}, and gives us a new perspective on the mass of Komar in Section \ref{s3d}. Conclusions follow in Section \ref{s4}, along with appendices addressing gauge theory methods in geometric algebra.
\par
Most of the notation is introduced as it arises, but throughout we will use the geometrised units $c=G=1$ so that $\kappa=8\pi$ and the signature $\eta_{ab}=\mathrm{diag}(+,-,-,-)$.
\section{Energy-momentum and mass in general relativity}\label{s2}
\subsection{Previous work on the flat background}\label{s2a}

We begin by providing a terse introduction to the formalism of \onlinecite{butcher2}.
This work is grounded in the mapping $\phi:\mathcal{M}\to \check{\mathcal{M}}$ from the physical spacetime, $\mathcal{M}$, with metric $g_{ab}$ to a flat background, $\check{\mathcal{M}}$, with metric $\check{g}_{ab}$. Here, Roman letters label Penrose's abstract indices\footnote{See for example Wald's ubiquitous use of these in \onlinecite{wald}.} which may appear in either manifold. The background is furnished with four Lorentzian coordinate functions, $\{x^\mu\}$, labelled by Greek indices, so that in $\check{\mathcal{M}}$ the vector $\partial/\partial x^\mu$ has components $\check{e}_\mu^{\ a}$ which obey $\check{e}_\mu^{\ a}\check{e}_\nu^{\ b}\check{g}_{ab}=\eta_{\mu\nu}$ and $\check{\nabla}_a \check{e}_\mu^{\ b}=0$. In $\mathcal{M}$ the image coordinate functions $\{y^\mu\}$ are formed by the pullback of the $x^\mu$ at any $p\in \mathcal{M}$: $y^\mu(p)=\phi_*(x^\mu)(p)=x^\mu\circ\phi(p)$. From these image coordinates a new basis $\partial/\partial y^\mu$ has components $\left(\phi^*e_\mu\right)^{a}=\check{e}_\mu^{\ a}$. Crucially, whilst the $\check{e}_\mu^{\ a}$ are trivially components of Killing vectors in $\check{\mathcal{M}}$, the same is not generally true of the $e_\mu^{\ a}$ in $\mathcal{M}$ because of the presence of the gravitational field. In the physical spacetime this leads to the non-conservation of the four local four-current densities of matter energy-momentum, $J_\mu^{\ a}$, which are formed by contracting $T_{pq}$ with the new basis vectors
\begin{eqnarray}\label{noncons}
	\nabla_a J_\mu^{\ a}=T^a_b \nabla_a e_\mu^{\ b}\neq 0.
\end{eqnarray}
According to the most conventional perturbation scheme
 \begin{equation}\label{pert}
	\begin{aligned}
		\phi^* g_{a}=&\check{g}_{ab}+h_{ab}, \quad \phi^* g^{ab}=\check{g}^{ab}-h^{ab}+\mathcal{O}(h^2),
\end{aligned}
\end{equation}
the method of \onlinecite{butcher2} is to cancel this matter stress-energy `leak' as it is manifest in a flat background with the equal and opposite covariant divergence of some tensor, $\tau\sim\check{\nabla}h\check{\nabla}h$. This tensor is thus determined by the background relation
\begin{eqnarray}\label{motivator1}
	\check{\nabla}^{a}\tau_{aq}=-\left[\phi^{*}\left( T^{a}_{b}\nabla_{a}e_{\mu}^{\ b} \right)\check{e}^{\mu}_{\ q}\right]^{(2)},
\end{eqnarray}
where objects to $n$th order in $h$ will be identified with a parenthesised superscript. Remarkably, a symmetric superpotential-free ansatz for $\tau_{pq}$ combined with the harmonic gauge constraint,
\begin{eqnarray}
  \check{\nabla}^a\bar{h}_{ab} =0,\quad \bar{h}_{ab}=h_{ab}-\tfrac{1}{2}\eta_{ab}h,	
\end{eqnarray}
was found to yield a unique solution to \eqref{motivator1}:
\begin{eqnarray}\label{butcher_tensor}
  \kappa\bar{\tau}_{pq}=\tfrac{1}{4}\check{\nabla}_p h_{ab} \check{\nabla}_q \bar{h}^{ab}.
\end{eqnarray}
Note that the overbar notation when dealing with tensors signifies the trace-reverse, and has quite a different meaning in geometric algebra. 
\subsection{A non-linear generalisation}\label{s2b}

As we established above, the tensor $\tau_{pq}$ lives in $\check{\mathcal{M}}$, and we would like to augment it with higher order corrections in $h$. We can invent a new tensor for the full series
\begin{eqnarray}\label{series}
	\check{\mathscr{T}}_{pq}=\sum_{n=2}^{\infty}\check{\mathscr{T}}^{(n)}_{\ \ \ pq}\stackrel{?}{=}\tau_{pq}+\sum_{n=3}^{\infty}\check{\mathscr{T}}^{(n)}_{\ \ \ pq},
\end{eqnarray}
anticipating $\check{\mathscr{T}}_{pq}$ to be the pushforwards of some $\mathscr{T}_{pq}$ in $\mathcal{M}$. A natural extension of the theory to third perturbative order is to introduce the ansatz $\check{\mathscr{T}}^{(3)}\sim h \check{\nabla}h\check{\nabla}h$ to the equation
\begin{equation}\label{motivator1a}
	\begin{aligned}
		\check{\nabla}^{a}\check{\mathscr{T}}^{(3)}_{\ \ \ aq}\stackrel{?}{=}-\left[\phi^{*}\left( T^{a}_{b}\nabla_{a}e_{\mu}^{\ b} \right)\check{e}^{\mu}_{\ q}\right]^{(3)},
\end{aligned}
\end{equation}
however this has no solution under \eqref{pert} with or without the harmonic gauge condition\footnote{Neither this nor the asymmetry of the solution to \eqref{motivator1b} will be alleviated by the `central expansion' mentioned in \onlinecite{butcher4}.}. One way to proceed is to generalize the form of the background covariant derivative on  the LHS of \eqref{motivator1a} by introducing some `friction connection' of the form $F^{(1)}\sim \check{\nabla}h$ which couples to $\tau_{pq}$. This connection will be constructed so as to account for the apparently ineradicable non-conservation as it appears even in $\check{\mathcal{M}}$. Accordingly, the third-order correction must instead obey 
\begin{equation}\label{motivator1b}
	\begin{aligned}
		\check{\nabla}^{a}\check{\mathscr{T}}^{(3)}_{\ \ \ aq}-\check{g}^{ap}\big(F^{(1)c}_{\ \ \ \ pa}\tau_{cq}+F^{(1)c}_{\ \ \ \ qa}\tau_{pc}\big)\\=-\left[\phi^{*}\left( T^{a}_{b}\nabla_{a}e_{\mu}^{\ b} \right)\check{e}^{\mu}_{\ q}\right]^{(3)}.
\end{aligned}
\end{equation}
This turns out to be very fruitful. The ansatz for $\check{\mathscr{T}}^{(3)}_{\ \ \ pq}$ cannot be solved uniquely, but it gives a space of \textit{asymmetric} third-order corrections to $\tau_{pq}$. More importantly, by repeating the procedure at higher orders it quickly becomes apparent that the $F^{(n)a}_{\ \ \ \ bc}$ are in fact terms from the perturbative expansion in $h$ of the Levi-Civita connection, familiar in the physical spacetime as $\Gamma^a_{ab}$ or in torsion-free general relativity as the Christoffel symbols, $\christoffel{a}{bc}$. Things now become clearer: the apparent `friction' in the background is nothing more than curvature creeping into the theory at higher orders. Because the Levi-Civita connection is a function of the gradient of the metric, it makes its first appearance in the third-order equation, \eqref{motivator1b}, spoiling the flat-space picture as it does so: the quadratic $\tau_{pq}$ is a special case that makes \onlinecite{butcher2} possible. 
\par
It is now easy to extend the theory to all orders. We aim to soak up the matter stress-energy leak directly in the physical spacetime with the equal and opposite covariant divergence of some \textit{gravitational} stress-energy four-currents, $\mathscr{J}_\mu^{\ a}$. We will suppose these currents to be formed from some tensor $\mathscr{T}^a_{\ b}e_\mu^{\ b}=\mathscr{J}_{\mu}^{\ a}$, to be identified as a gravitational stress-energy tensor. Note in particular that so long as $\mathscr{T}_{pq}$ sports Penrose indices we really do mean it to be a tensor-valued object, which may be covariantly differentiated to give
\begin{equation}\label{premotivator2}
	\begin{aligned}
		\nabla_a \mathscr{J}_\mu^{\ a}&=\nabla_a \left(\mathscr{T}^a_{\ b}e_\mu^{\ b}\right)\\&=-\nabla_a J_\mu^{\ a}=-\tfrac{1}{\kappa}\nabla_a \left(G_b^a e_\mu^{\ b}\right).
\end{aligned}
\end{equation}
In the last equality, the matter stress-energy tensor is translated into curved spacetime using the Einstein equations. In general, the only such tensor that satisfies \eqref{premotivator2} is proportional to the Einstein tensor itself\footnote{In Section \ref{s3a} we will explore the gauge theory version of this tautology.}. To distance ourselves from this fact, notice that \eqref{premotivator2} takes on a tidier form in the $\{y^\mu\}$ coordinate system. In this case, the components of the basis are given by the Kronecker delta, so \eqref{premotivator2} reduces to
\begin{eqnarray}\label{motivator3}
	\kappa\left(\partial_\alpha \mathscr{T}^\alpha_{\ \lambda}+\christoffel{\alpha}{\beta\alpha}\mathscr{T}^\beta_{\ \lambda}\right)=-G_\beta^\alpha \christoffel{\beta}{\alpha\lambda}.
\end{eqnarray}
We want $\mathscr{T}_{\mu\nu}$ to be second order in the first derivatives of the metric. Using therefore the ansatz $\mathscr{T}\sim\partial g \partial g$ it can be shown that \eqref{motivator3} has a \textit{unique}\footnote{The calculation is longer than that required to obtain $\tau_{ab}$ in \onlinecite{butcher2}, but takes a similar form.} solution without the need for further gauge constraints (such as the harmonic coordinate condition). It may be written compactly in trace-reversed form as the following function of spacetime:
\begin{equation}\label{object}
\begin{aligned}
	&\kappa\bar{\mathscr{T}}_{\sigma\lambda}=\tfrac{1}{4}g^{\beta\gamma}g^{\varepsilon\alpha}\big( \partial_\sigma g_{\varepsilon\beta}\partial_\lambda g_{\alpha\gamma}-\partial_\sigma g_{\varepsilon\alpha}\partial_\lambda g_{\beta\gamma}\\&+\partial_\alpha g_{\sigma\varepsilon}\partial_\lambda g_{\beta\gamma}+\partial_\alpha g_{\beta\gamma}\partial_\lambda g_{\sigma\varepsilon}-2\partial_\beta g_{\sigma\varepsilon}\partial_\lambda g_{\alpha\gamma}\big).
\end{aligned}
\end{equation}
\subsection{Einstein's pseudotensor}\label{s2c}

Two sinister features of the function \eqref{object} are immediately obvious: firstly it is asymmetric in its indices and secondly it emphatically does \textit{not} constitute a tensor definition. These features are explained by a third observation, that \eqref{object} are identically the components of the transposed\footnote{Because of the natural definition of gravitational energy-momentum currents in \onlinecite{butcher2}, the extension $\mathscr{T}_{ab}$ is defined with a transpose relative to the pseudotensor of Einstein.} Einstein pseudotensor, $_Et_{a}^{\  b}$, in the $\left\{ y^\mu \right\}$ coordinate system
\begin{eqnarray}
  \mathscr{T}_{\sigma\lambda}={_Et_{\lambda\sigma}}.
\end{eqnarray}
\par
In hindsight it is easy to see why we have arrived at the oldest description of gravitational energetics in general relativity. We mentioned in Section \ref{s1} that Einstein's energy-momentum complex admits a special conservation law. Given the partitioning in \eqref{einstein_comp} this law becomes 
\begin{eqnarray}\label{damnation}
  \partial_a \big(\sqrt{-g} _Et_{b}^{\ a}\big)=-\partial_a\big( \sqrt{-g}T_b^a\big).
\end{eqnarray}
If we cast \eqref{damnation} in the $\left\{ y^\mu \right\}$ coordinates and differentiate the metric determinant according to
\begin{eqnarray}
	\partial_\alpha \sqrt{-g}=-\tfrac{1}{2}\sqrt{-g}g^{\mu\nu}\partial_\alpha g_{\mu\nu}=\sqrt{-g}\christoffel{\mu}{\alpha\mu},	
\end{eqnarray}
we are left (once the indices on $\mathscr{T}_{\alpha\beta}$ are swapped) with precisely the motivating equation \eqref{motivator3}, which is the generalisation of the local conservation law of \onlinecite{butcher2}. Whilst $_Et_{pq}$ is the same quadratic function of the metric derivatives in all coordinate systems (and hence is a pseudotensor), $\mathscr{T}_{pq}$ was set up as a tensor that coincides with $_Et_{pq}$ in the $\left\{ y^\mu \right\}$ coordinate system. As with the pseudotensor of Landau and Lifshitz, it is possible to write $_Et_{\alpha\beta}$ as a quadratic function of the $\christoffel{\alpha}{\beta\gamma}$, so $\mathscr{T}_{pq}$ can be constructed as a quadratic function of the $\check{\nabla}_b e_\mu^{\ a}$ and $\check{\nabla}_b e^\mu_{\ a}$, using the $e_\mu^{\ a}$ and $e^\mu_{\ a}$ to contract away all Lorentz indices. This does not of course constitute progress, because we have simply recast a pseudotensor as a tensor-valued function of a privileged coordinate system.
\par 
In the harmonic gauge and to lowest order, $\check{\mathscr{T}}_{pq}$ and $\tau_{pq}$ do not agree, since
\begin{equation}\label{difference}
\begin{aligned}
	\kappa\bar{\check{\mathscr{T}}}^{(2)}_{\ \ \ pq}=&\tfrac{1}{4}\check{\nabla}_p h_{ab} \check{\nabla}_q \bar{h}^{ab}+\tfrac{1}{4} \check{\nabla}_a h \check{\nabla}_q h^{a}_{p}\\&-\tfrac{1}{2} \check{\nabla}_a h_{pb} \check{\nabla}_q h^{ab},
\end{aligned}
\end{equation}
which differs from $\kappa \bar{\tau}_{pq}$ by the last two terms\footnote{So, the series \eqref{series} was only nearly correct.}. There is no contradiction here because so long as the harmonic condition holds it can readily be shown that
\begin{eqnarray}
	\check{\nabla}^a \big( \check{\mathscr{T}}^{(2)}_{\ \ \ aq}-\tau_{aq} \big)=0.
\end{eqnarray}
Hence, the tensor $\tau_{pq}$ is formed by trimming an identically conserved quantity (a `gauge current') from the linearised Einstein pseudotensor in the harmonic gauge.

\subsection{The linearised Klein-Gordon correspondence}\label{s2d}

The tensor $\tau_{ab}$ lends itself well to the Newtonian limit of gravitostatics. An inertial observer with velocity $v^\mu$ near a perfect fluid in hydrostatic equilibrium finds the matter stress-energy tensor to be $\check{T}_{\mu\nu}=\rho v_\mu v_\nu$ to lowest order in $h$ (which is to say they can neglect pressure). The linearised field equations in the harmonic gauge
\begin{eqnarray}
  \check{\nabla}^2 \bar{h}_{ab}=-2\kappa \check{T}_{ab},
\end{eqnarray}
yield the familiar Newtonian potential
\begin{eqnarray}
  h_{\mu\nu}=2\varphi(2v_\mu v_\nu-\eta_{\nu\mu}),
\end{eqnarray}
which obeys
\begin{eqnarray}\label{field_equation}
	\check{\nabla}^2\varphi=-\kappa\rho/2. 
\end{eqnarray}
To give a minimal example, a compact spherically symmetric distribution of total mass $M_{\mathrm{T}}$ gives rise to the external potential $\varphi=-M_{\mathrm{T}}/r$ and metric perturbation
\begin{eqnarray}\label{weak_isotropic}
	\mathrm{d}s^2=\left( 1-2M_{\mathrm{T}}G/r \right)\mathrm{d}t^2-\left( 1+2M_{\mathrm{T}}G/r \right)\mathrm{d}x_i^2
\end{eqnarray}
where $\mathrm{d}x_i^2=\mathrm{d}x_i\mathrm{d}x_i$ for $i\in \left\{ 1,2,3 \right\}$. This is the Newtonian limit of the \textit{rectangular isotropic} line element for Schwarzschild spacetime: we therefore see that isotropic coordinates arise naturally in linear gravitostatics.
\par
\begin{figure}[htp]
  \includegraphics[width=\linewidth]{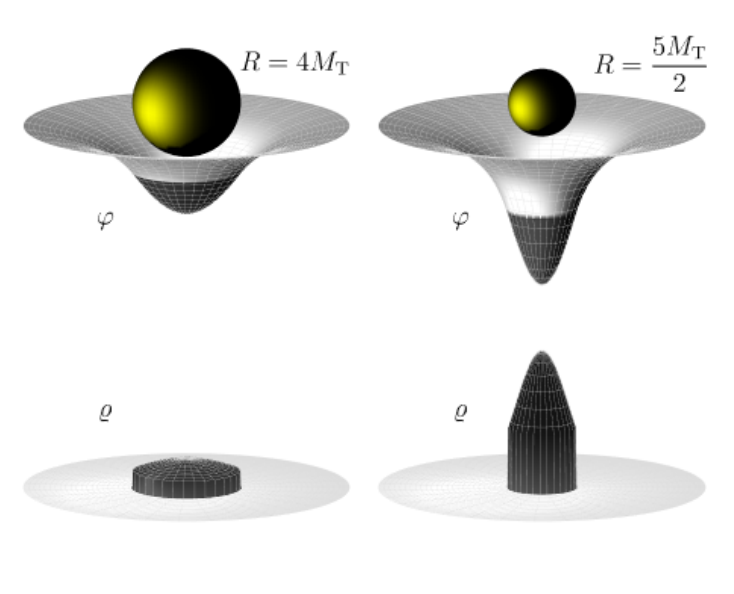}
  \caption{\label{figure-1} The picture of energetics to be developed in Section \ref{s3c} using isotropic coordinates in the gauge theory approach. The pseudotensors of M{\o}ller and Einstein both describe gravitational stress-energy as if the gravitational potential, $\varphi$, were a real Klein-Gordon field, generated by a source density, $\varrho$, which in turn integrates to give the gravitational mass, $M_T$, of the system. Here, for a pair of highly relativistic Schwarzschild stars approaching collapse to a black hole, the Newtonian form of the potential, $\varphi=-M_T/r$, is preserved right down to the stellar surface. We have already seen in Section \ref{s2d} how this `Klein-Gordon correspondence' is reflected by the tensor of Butcher in the Newtonian limit.}
\end{figure}

An apparently unrelated way (see \onlinecite{butcher2}) to arrive at the field equation \eqref{field_equation} is through the Klein-Gordon theory\footnote{The Klein-Gordon field here is real, but for convenience the kinetic term lacks a factor of $1/2$, in common with the complex (charged) theory.}
\begin{eqnarray}\label{KG_lagrangian1}
  _{KG}\mathcal{L}=\tfrac{1}{\kappa}\partial_\alpha \varphi \partial^\alpha \varphi-\varphi\rho,
\end{eqnarray}
which highlights a very curious feature of $\tau_{ab}$. Since 
\begin{eqnarray}
	\kappa\bar{\tau}_{\mu\nu}=2\partial_\mu\varphi\partial_\nu\varphi,
\end{eqnarray}
we see that $\tau_{\mu\nu}$ is describing the stress and energy bound up in the Newtonian potential as if it were a scalar field generated by matter. At linear order there is of course room for this `Klein-Gordon correspondence' to appear coincidental, but equipped with the generalisation to Einstein's pseudotensor, we will show in Section \ref{s3c} that the principle does in fact apply at all orders -- this is illustrated in Figure \ref{figure-1} below for a pair of Schwarzschild stars with the same gravitational mass but different densities. To make this generalisation, we will need not only isotropic coordinates, but some notion of the flat background, $\check{\mathcal{M}}$, in full gravity. Such a construct is provided naturally by the gauge theory approach.
\subsection{Mass in general relativity}\label{s2e}

\begin{figure}[htp]
  \includegraphics[width=\linewidth]{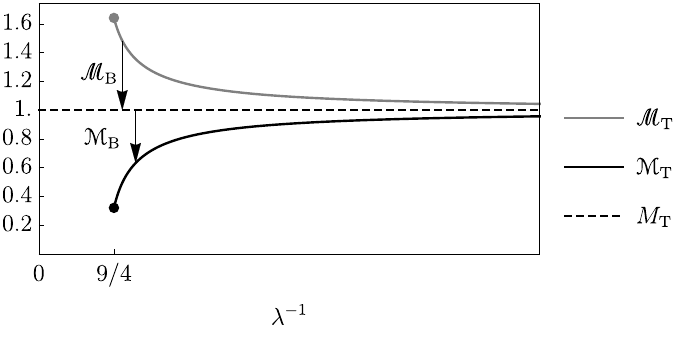}
  \caption{\label{figure-2} Relativistic binding energies of the Schwarzschild star. The proper mass, $\mathproper{M}_T$ is greater than the gravitational mass, $M_T$, which is greater than the conserved mass $\mathscr{M}_T$. The relativistic masses diverge as the Newtonian parameter $\lambda^{-1}=\bar{R}/M_T$ shrinks. The Schwarzschild star is notoriously unstable for $\lambda^{-1}\leq 9/4$ in Schwarzschild coordinates -- though according to Buchdahl's theorem \cite{buchdahl} such stars are the most compact that can ever form.}
\end{figure}

Before addressing energy localisation in the gauge theory approach, we will make some observations regarding relativistic mass. The spacetimes of particular interest to us will be not only stationary but static. Consequently there will be a global timelike Killing vector $K^a$, such that if a Cauchy surface $\Sigma_t$ were defined to be a contour of the Killing parameter $t$, $K^a$ would be orthogonal to that surface. Furthermore the spacetimes will be asymptotically flat in the sense discussed above, spherically symmetric\footnote{However we will not make explicit use of the trio of Killing vectors associated with the spherical symmetry.} and regular everywhere. Of course, this restricts us to precisely those spacetimes which were of earliest astrophysical interest, since they accommodate the unspun relativistic stars. Of such stars, we will consider only those composed of a perfect fluid. It is very convenient indeed to study these systems using \textit{Schwarzschild-like} coordinates, with general line element
\begin{eqnarray}\label{schwarzschild_line_element}
  \mathrm{d}s^2=e^{A}\mathrm{d}t^2-e^{B}\mathrm{d}\bar{r}^2-\bar{r}^2(\mathrm{d}\theta^2+\sin^2\theta\mathrm{d}\phi^2).
\end{eqnarray}
Such coordinates have the advantage of preserving the ratio of $2\pi$ between the radial coordinate $\bar{r}$ and the proper distance about the equator. Less frequently used (we have met them already in Section \ref{s2d} and shall use them extensively in Section \ref{s3c}) are the \textit{isotropic} coordinates,
\begin{eqnarray}\label{isotropic_line_element}
  \mathrm{d}s^2=e^{A}\mathrm{d}t^2-e^{C}[\mathrm{d}r^2-r^2(\mathrm{d}\theta^2+\sin^2\theta\mathrm{d}\phi^2)].
\end{eqnarray}
The stress-energy tensor of a perfect fluid with proper density $\rho$, pressure $P$ and bulk four-velocity $u^a$ is
\begin{eqnarray}\label{perfect_fluid_set}
  T^{ab}=\left( \rho+P \right)u^a u^b-Pg^{ab}.
\end{eqnarray}
Of the Einstein equations, we will particularly require
\begin{eqnarray}\label{einstein_equations}
  \begin{aligned}
    \kappa\rho&=B'e^{-B}/\bar{r}+(1-e^{-B})/\bar{r}^2,\\
    \kappa P&=A' e^{-B}/\bar{r}-(1-e^{-B})/\bar{r}^2,
  \end{aligned}
\end{eqnarray}
where throughout this section, prime denotes differentiation with respect to $\bar{r}$. Some very useful derived results are then
\begin{align}\label{derived_1}
  A'=\frac{2M\left( 1+4\pi \bar{r}^3P/M \right)}{\bar{r}^2\left( 1-2M/\bar{r} \right)},
\end{align}
and
\begin{eqnarray}\label{derived_2}
  A'=-2P'/\left( \rho+P \right).
\end{eqnarray}
From \eqref{derived_1} and \eqref{derived_2} we can assemble the famous Tolman-Oppenheimer-Volkoff equation
\begin{eqnarray}\label{tov}
  P'=-\frac{M\left( \rho+P \right)\left( 1+4\pi \bar{r}^3P/M \right)}{\bar{r}^2\left( 1-2M/\bar{r} \right)},
\end{eqnarray}
which is used to construct solutions for relativistic stars. The first equation in \eqref{einstein_equations} can be written as
\begin{eqnarray}
  \left[\bar{r}\left(1-e^{-B}\right)\right]'=8\pi\bar{r}^2\rho.
\end{eqnarray}
The only integral that guarantees a regular metric at the origin is
\begin{eqnarray}
	e^{-B}=1-2M/\bar{r}
\end{eqnarray}
where we have introduced our first mass function
\begin{eqnarray}\label{mass_function_schwarzschild}
  M=\int_0^{\bar{r}} \mathrm{d}\tilde{\bar{r}}\,4\pi {\tilde{\bar{r}}}^2\rho.
\end{eqnarray}
As is pointed out in \onlinecite{buchdahl}, the mass defined by \eqref{mass_function_schwarzschild} does not correspond to any invariant quantity whatever, and serves a convenient but potentially misleading `book-keeping' purpose in Schwarzschild-like coordinates. At the \textit{surface} of the fluid, $\bar{r}=\bar{R}$, where we are obliged to glue $e^B$ to the Schwarzschild volume element, we find $M({\bar{R}})=M_{T}$ where $M_{T}$ is the `total gravitational mass'. We established already in the introduction that $M_T$ is a good physical quantity in these systems, and may be recovered through the methods of Komar and Bondi or Arnowitt, Deser and Misner.  
The second mass we will consider has a clearer physical motivation at arbitrary radius. It is the integral of the matter density over the proper volume,
\begin{eqnarray}\label{mass_function_proper}
  \mathproper{M}=\int_0^{\bar{r}} \mathrm{d}\tilde{\bar{r}}\,4\pi {\tilde{\bar{r}}}^2\rho e^{B/2}.
\end{eqnarray}
The quantity $\mathproper{M}_{T}$ is known as the \textit{proper mass} of the fluid. The proper and gravitational masses are related through a quantity $\mathproper{M}_{{B}}$ which is traditionally taken to be the gravitational binding energy
\begin{eqnarray}\label{proper_binding_energy}
	\mathproper{M}_{T}=M_{T}+\mathproper{M}_{{B}}.
\end{eqnarray}
Let us now use the line element \eqref{schwarzschild_line_element} to introduce a further mass function,
\begin{eqnarray}\label{flat_mass_function}
  \mathscr{M}=\int_0^{\bar{r}} \mathrm{d}\tilde{\bar{r}}\,4\pi {\tilde{\bar{r}}}^2\rho e^{A/2+B/2},
\end{eqnarray}
which, up to a normalisation of $K^a$ corresponds to the quantity mentioned in \eqref{consone}. In some sense $\mathscr{M}$ is `complimentary' to $\mathproper{M}$, in that it allows us to define an alternative binding energy
\begin{eqnarray}\label{flat_binding_energy}
	\mathscr{M}_{T}=M_{T}-\mathscr{M}_{{B}}.
\end{eqnarray}
The choice of signs in \eqref{proper_binding_energy} and \eqref{flat_binding_energy} is to reflect the fact that binding energy so defined should be positive in order for the star to be stable: the behaviour of the mass functions $\mathproper{M}_T$ and $\mathscr{M}_T$ is compared in Figure \ref{figure-2} for the Schwarzschild star of gravitational mass $M_T$ at various degrees of gravitational collapse. Having introduced $\mathscr{M}$, we notice that the line element \eqref{schwarzschild_line_element} suggests a second quantity which integrates to $M_T$. The first step is to expand \eqref{flat_mass_function} by parts
\begin{eqnarray}\label{parts}
  \mathscr{M}_{T}=M_{T}-\int_0^{\bar{R}} \mathrm{d}{\tilde{\bar{r}}}\,  M\left( A'+B' \right)e^{A/2+B/2}/2.
\end{eqnarray}
Then, we can apply relations \eqref{derived_1} and \eqref{derived_2} to show
\begin{eqnarray}
  \begin{aligned}
    &\int_0^{\bar{R}}\mathrm{d}\tilde{\bar{r}}12\pi\tilde{\bar{r}}^2 Pe^{A/2+B/2}\\&=-\int_0^{\bar{R}}\mathrm{d}\tilde{\bar{r}}4\pi\tilde{\bar{r}}^3(P'+P(A'+B')/2)e^{A/2+B/2}\\
    &=\int_0^{\bar{R}}\mathrm{d}\tilde{\bar{r}}4\pi\tilde{\bar{r}}^3(\rho A'-PB')e^{A/2+B/2}/2.
\end{aligned}
\end{eqnarray}
From here, by inserting the two field equations of \eqref{einstein_equations} and comparing with \eqref{parts} we see that if a mass function is defined by
\begin{eqnarray}\label{komar_mass_function}
  \mathfrak{M}=\int_0^{\bar{r}}\mathrm{d}\tilde{\bar{r}}4\pi\tilde{\bar{r}}(\rho+3P)e^{A/2+B/2},
\end{eqnarray}
we will have agreement with the gravitational mass at the stellar surface
\begin{eqnarray}
  \mathfrak{M}_T=M_T.
\end{eqnarray}
The formula \eqref{komar_mass_function} for $\mathfrak{M}_T$ corresponds to a very powerful definition of gravitational mass in stationary, asymptotically flat spacetimes known as the \textit{Komar mass}. We will briefly outline the physical motivation behind this quantity, bearing in mind that the techniques used in the derivation -- ubiquitous in general relativity -- will later need to be imported into the gauge theory of Section \ref{s3}. 
If a unit test mass is suspended above the star, so that it has four-velocity $u^a=K^a/K$, the force applied at spatial infinity to keep it there is
\begin{eqnarray}
  F_a=u^b\nabla_{b}K_a.
\end{eqnarray}
If such a mass is distributed over closed 2-surface $\partial V$ which contains the star, the observer at infinity must apply an outward force
\begin{eqnarray}\label{total_force}
  F=\oint_{\partial V} |\mathrm{d}^2x| n^a u^b \nabla_b K_a,
\end{eqnarray}
where $n^a$ is the unit normal to $\partial V$ and $|\mathrm{d}^2x|$ is the scalar volume element on $\partial V$. Of course, the static condition ensures $n^au_a=0$. Now as $\partial V$ is retracted to the asymptotically flat region at spatial infinity, this force may be unambiguously equated with the gravitational mass, and accordingly the Komar mass is defined $\kappa F =2\mathfrak{M}_T$. An application of the Killing equation to \eqref{total_force} allows the Komar mass associated with $V$ to be written in the powerful language of differential forms
\begin{eqnarray}\label{komar_surface_integral}
  \mathfrak{M}=-\tfrac{1}{\kappa}\oint_{\partial V}\epsilon_{abcd}\nabla^c K^d=\oint_{\partial V}\alpha_{ab}.
\end{eqnarray}
In \eqref{komar_surface_integral}, $\epsilon_{abcd}$ is the natural volume element on $\mathcal{M}$ imposed by $g_{ab}$ and $\alpha_{ab}$ is the resultant two-form to be integrated over $\partial V$. Use of the metric to define volume elements carries profound advantage when applying Stokes' theorem:
\begin{eqnarray}\label{stokes}
  \oint_{\partial V}\alpha_{ab}=\int_V d\alpha_{abc}.
\end{eqnarray}
In the absence of torsion, the operation $d$ which generates an $n+1$-form from an $n$-form is independent of the derivative operator used to perform it, allowing for the natural choice, $\nabla_a$. Since the covariant derivative of the natural volume element vanishes, \eqref{stokes} can be used to write \eqref{komar_surface_integral} as a volume integral over the second covariant derivative of $K^a$ -- in this form \eqref{stokes} is known as \textit{Gauss' theorem}. The well known relation for Killing vectors\footnote{We re-derive this using gauge theory methods in Appendix \ref{a3}.}
\begin{eqnarray}
  \nabla_a\nabla^aK^b=-R^b_aK^a,
\end{eqnarray}
then enables the Komar mass to be written
\begin{eqnarray}\label{komar_volume}
  \mathfrak{M}_T=\tfrac{2}{\kappa}\int_V|\mathrm{d}^3x| R_{ab}u^a K^b,
\end{eqnarray}
where $|\mathrm{d}^3x|$ is the scalar volume element on $V$. 
By exposing the Ricci tensor in the integrand, we see that the contribution from the vacuum vanishes, and so the two-surface may be arbitrarily deformed around the star it encloses. To arrive at \eqref{komar_volume}, the Ricci tensor (i.e. a gravitational quantity) was extracted through an integral theorem, reminding us of the ultimate connection between geometry and gravity in general relativity. We will later see how to reproduce the Komar mass using the gauge theory approach, which eschews such a connection. 
By applying the Einstein equations to the perfect fluid, \eqref{perfect_fluid_set}, it is easy to see how \eqref{komar_volume} is equivalent to the formula given in \eqref{komar_mass_function}.
\par
In the Newtonian limit we can expect $\bar{r}/M$ always to be large within the star, and to be of the same order as $\rho/P$ and the Newtonian parameter $\lambda^{-1}=\bar{R}/M_T$. In the same limit, the gravitational force which binds the perfect fluid is expected to follow an $r^{-1}$ potential. Expanding the total proper mass according to
\begin{eqnarray}
  \mathproper{M}_T=\sum_{n=1}^\infty {\mathproper{M}_T}_n\lambda^n.
\end{eqnarray}
and substituting with \eqref{tov} we find
\begin{equation}
  \begin{aligned}
    \mathproper{M}_{T}&=M_{T}+\int_0^{\bar{R}} \mathrm{d}\tilde{\bar{r}}\,4\pi\tilde{\bar{r}}M\rho +\mathcal{O}(\lambda^3)\\
    &=M_{T}-\int_0^{\bar{R}} \mathrm{d}\tilde{\bar{r}}\,4\pi\tilde{\bar{r}}^3P' +\mathcal{O}(\lambda^3).
\end{aligned}
\end{equation}
If we consider the star to be made up of an ideal gas, a final application of integration by parts shows that the quantity $\mathproper{M}_{{B}}$ is equivalent in the Newtonian limit to twice the internal kinetic energy
 \begin{eqnarray}\label{schwarzschild_proper_virial}
   \mathproper{M}_{T}= M_{T}+\int_0^{\bar{R}} \mathrm{d}\tilde{\bar{r}}\, 12\pi{\tilde{\bar{r}}}^2P+\mathcal{O}(\lambda^3),
\end{eqnarray}
which is a statement of the virial theorem. 
\par
Conversely, it is possible to do exactly the same thing with the quantity defined by \eqref{flat_mass_function}. $\mathscr{M}_{T}$ can be related to $M_{T}$ by an application of integration by parts. Using \eqref{derived_2} and \eqref{tov} this produces 
\begin{equation}\label{virial_expansion}
	\begin{aligned}
	  \mathscr{M}_{T}&=M_{T}-\int_0^{\bar{R}}\mathrm{d}{\tilde{\bar{r}}}\,  M\left( A'+B' \right)e^{A/2+B/2}/2\\
	  &=M_{T}-\int_0^{\bar{R}} \mathrm{d}{\tilde{\bar{r}}}\,  4\pi \tilde{\bar{r}}M\left( \rho+P \right)e^{A/2+3B/2}\\
	  &=M_{T}+\int_0^{\bar{R}} \mathrm{d}{\tilde{\bar{r}}}\,   8\pi \tilde{\bar{r}}MP'e^{A/2+3B/2}/A '.
	\end{aligned}
\end{equation}
If we now expand \eqref{virial_expansion} in the Newtonian limit we will find
\begin{eqnarray}
	e^{A/2+3B/2}/A'=\bar{r}^2/2M+\mathcal{O}\left(\lambda\right).
\end{eqnarray}
Hence as before, a second application of integration by parts gives a complementary statement of the virial theorem
\begin{eqnarray}\label{schwarzschild_flat_virial}
  \mathscr{M}_{T}=M_{T}-\int_0^{\tilde{\bar{R}}} \mathrm{d}{\tilde{\bar{r}}}\,  12\pi{\tilde{\bar{r}}}^2P+\mathcal{O}(\lambda^3).
\end{eqnarray}
Having introduced the quantities $\mathfrak{M}$ and $\mathscr{M}$, the necessity of integration by parts in obtaining \eqref{schwarzschild_proper_virial} and \eqref{schwarzschild_flat_virial} might be understood as a consequence of relying on the coordinate-dependent quantity, $M$, to define gravitational mass. An immediately striking feature of \eqref{komar_mass_function} in the context of the virial theorem is the factor of $\rho+3P$ in the integrand. Following this suggestive factor, we can write an alternative to \eqref{schwarzschild_flat_virial} which expresses the virial theorem as a \textit{local} property of the perfect fluid, i.e. true at all radii
\begin{eqnarray}\label{komar_virial}
  \mathscr{M}=\mathfrak{M}-\int_0^{\tilde{\bar{r}}} \mathrm{d}{\tilde{\bar{r}}}\,  12\pi{\tilde{\bar{r}}}^2P+\mathcal{O}(\lambda^3).
\end{eqnarray}
Of course, interpretation of \eqref{schwarzschild_flat_virial} or \eqref{komar_virial} as statements of the virial therem rely entirely on the interpretation of $\mathscr{M}_B$ as a binding energy in \eqref{flat_binding_energy}. Whilst this interpretation is not immediately suggested by \eqref{consone}, it begins to look less arbitrary when we attempt to localise gravitational energy in gauge theory gravity. 
\section{The view from gauge theory gravity}\label{s3}
In Section \ref{s2a} we employed a popular but nearly unnecessary mathematical procedure for linearising gravity by moving between two manifolds.
Nearly, because whilst it is quite feasible to treat linearisation as an algebraic problem without ever leaving the physical spacetime, the introduction of a flat background could be considered suggestive of the gauge theory approach. 
On the one hand, $\mathcal{M}$ is taken to contain some interesting geometry (and hence gravity) imposed by $g_{ab}$. 
On the other, the geometry of the flat background $\check{\mathcal{M}}$ is entirely trivial, $\check{g}_{ab}$ imparting it with nothing more than the Minkowkian signature. 
In $\check{\mathcal{M}}$ the geometry of $\mathcal{M}$ is represented by a collection of tensor fields which, being small, can be managed by a series expansion. 
This of course becomes either impractical or impossible in the general case of strong gravitational fields.
In gauge theories of gravity, the gravitational gauge fields do not dictate the geometry of the manifold which contains them, that being simply Minkowski space, $M_4$, nor yet need they be expressed through a series expansion. 
In the particular case of gauge theory gravity, the nomenclature is influenced by the use of the spacetime algebra, and $M_4$ is often referred to as the vector space, $\{x\}$. We will stop short of formally identifying $\check{\mathcal{M}}$ with $\{x\}$ because they belong to \textit{entirely} different theories. 
Many previous articles on gauge theory gravity have included their own substantial primers on the spacetime algebra. We will break with tradition by referring the reader to more dedicated sources, and outlining in name only the principles essential to the gauge theory approach, which are as follows. 
\par
The spacetime algebra is a graded algebra of multivectors spanned by one scalar, four vectors, six bivectors, four trivectors and one pseudoscalar. Particular grades of a multivector are extracted with subscripted chevrons, with the absence of a subscript indicating the scalar, or grade $0$, part. The fundamental operations are addition and the geometric product, which is denoted by a simple juxtaposition of variables. Particularly useful compositions of these operations are the interior, exterior, commutative and scalar products, respectively $\cdot$, $\wedge$, $\times$ and $*$. We may chose to work either with the orthonormal Lorentz-basis of vectors, $\gamma_\mu$, and dual basis, $\gamma^\mu$, or with arbitrary constant vectors, denoted by lowercase Latin letters such as $a$ and  its dual, $\partial_a$. Tensors of second rank are represented by vector-valued linear functions of such vectors, for example the Ricci tensor, $\mathcal{R}(a)$, or position gauge field, $\underline{\mathsf{h}}(a)$, which has the same function as the tetrad. Whenever a linear function has the same rank as its argument, the underbar-overbar notation is useful in distinguishing the function from its adjoint, which roughly corresponds to the commutation of tensor indices. A related concept is the inverse of the linear function, for example $\underline{\mathsf{h}}^{-1}\underline{\mathsf{h}}(a)=a$. The spin connection is a tensor of third rank, and correspondingly the rotation gauge field is a bivector-valued linear function, $\Omega(a)$. The derivative with respect to position in $\{x\}$ is simply $\nabla$, whilst the covariant derivative is $\mathcal{D}$. Overdot notation and arrows may be used in an obvious manner to indicate the intended target of a derivative operation, whenever nested parentheses on a single line cannot elegantly convey the mathematical content of an expression. For a working understanding of these techniques, we recommend either Part I of \onlinecite{lasenby} or Chapter 13 of \onlinecite{doran} as compact introductions to gauge theory gravity. In addition, reference to Chapter 1 of \onlinecite{hestenes2}, which presents many essential geometric algebra identities in short order, may be very beneficial. An alternative introduction to gauge theory gravity is to be found in \onlinecite{hestenes}, although it differs from our treatment in its emphasis on \textit{gravity frames}, $g_\mu=\underline{\mathsf{h}}^{-1}(\gamma_\mu)$ and $g^\mu=\bar{\mathsf{h}}(\gamma^\mu)$. Whilst such an approach is more reminiscent of differential geometry, we will instead try to take full advantage of the spacetime algebra by expressing relations in \textit{frame-free form} wherever possible.
\subsection{The Einstein-Hilbert Lagrangian}\label{s3a}
The total action of gauge theory gravity as defined in \onlinecite{lasenby} corresponds to that of Einstein and Hilbert
\begin{eqnarray}\label{gtg_action}
  \begin{aligned}
    S&=\int| \mathrm{d}^4 x | \left( \tfrac{1}{2}\mathcal{R}-\kappa\mathcal{L}_m \right)\det \underline{\mathsf{h}}^{-1}\\&=-\kappa\int |\mathrm{d}^4x|\left( \mathproper{L}_g+\mathproper{L}_m \right).
\end{aligned}
\end{eqnarray}
By extracting the conventional factor of $-\kappa$ we can consider gravitational and matter Lagrangian densities as scalar densities on $M_4$, for these we use a distinct script. This should not be confused with the convention in gauge theory gravity of using calligraphic script for quantities which are position gauge covariant
The gravitational Lagrangian density is therefore the density of the Ricci scalar
\begin{eqnarray}\label{grav_sect}
  \begin{alignedat}{2}
    \mathproper{L}_g &=&&-\tfrac{1}{2\kappa}\mathcal{R}\det\underline{\mathsf{h}}^{-1}\\
    &=&&\tfrac{1}{\kappa}\left( \partial_a \wedge \partial_b \right)\cdot \big[ \underline{\mathsf{h}}(a)\cdot\dot{\nabla}\dot{\Omega}(\underline{\mathsf{h}}(b))\\& &&+\tfrac{1}{2}\Omega(\underline{\mathsf{h}}(a))\times\Omega(\underline{\mathsf{h}}(b))  \big] \det \underline{\mathsf{h}}^{-1}.
\end{alignedat}
\end{eqnarray}
As with Einstein-Cartan theory, gauge theory gravity has two equations of motion. Variation with respect to the $\bar{\mathsf{h}}(a)$-field produces the Einstein equation,
\begin{eqnarray}\label{einstein_gtg}
  \mathcal{G}(a)=\kappa\mathcal{T}(a),
\end{eqnarray}
where the \textit{functional} or \textit{dynamical} stress-energy tensor of matter is defined
\begin{eqnarray}
  \mathcal{T}(\underline{\mathsf{h}}^{-1}(a))=\det \underline{\mathsf{h}}\,\partial_{\bar{\mathsf{h}}(a)}(\mathcal{L}_m\det\underline{\mathsf{h}}^{-1}).
\end{eqnarray}
Similarly, the spin-torsion equation is the equation of motion of the $\Omega(a)$-field
\begin{eqnarray}
  \mathcal{H}(a)=\kappa\mathcal{S}(a), \quad \mathcal{S}(\bar{\mathsf{h}}(a))=\partial_{\Omega(a)}\mathcal{L}_m,
\end{eqnarray}
so the gauge theory corresponding to general relativity follows from a matter Lagrangian in which the rotation gauge fields do not appear, e.g. 
\begin{eqnarray}\label{generic_mat}
  \mathcal{L}_m=\mathcal{L}_m (\mathcal{D}\phi_i, \phi_i),
\end{eqnarray}
where the $\phi_i$ are scalar matter fields.\footnote{An important exception to this rule is electromagnetism in a gravitational background, where the $\phi_i$ are replaced by the vector potential $\mathcal{A}$. This does not cause a problem because $\mathcal{D}\mathcal{A}$ does not appear directly in the Maxwell Lagrangian density, though it remains covariant.}
In this case, the vanishing of the torsion bivector, $\mathcal{H}(a)$, allows us to formulate the rotational gauge field in terms of the displacement gauge field
\begin{eqnarray}\label{torsion_one}
  \begin{aligned}
  \omega(b)=&\bar{\mathsf{h}}(\dot{\nabla})\wedge\dot{\bar{\mathsf{h}}}\bar{\mathsf{h}}^{-1}(b)\\
  &-\tfrac{1}{2}b\cdot\big[\partial_c\wedge\bar{\mathsf{h}}(\dot{\nabla})\wedge\dot{\bar{\mathsf{h}}}\bar{\mathsf{h}}^{-1}(c)\big],
\end{aligned}
\end{eqnarray}
where $\omega(a)=\Omega(\underline{\mathsf{h}}(a))$ is covariant under displacements, and in practice the following contraction will also be very useful
\begin{eqnarray}\label{torsion_two}
  \begin{aligned}
    \partial_b\cdot\omega(b)=&\dot{\bar{\mathsf{h}}}(\dot{\nabla})-\bar{\mathsf{h}}(\dot{\nabla})\partial_c\cdot\dot{\bar{\mathsf{h}}}\bar{\mathsf{h}}^{-1}(c).
\end{aligned}
\end{eqnarray}
The invariance of the total action \eqref{gtg_action} under global spacetime translations along some constant vector $n$ allows us to form the \textit{canonical} stress-energy tensor associated with that action 
\begin{eqnarray}\label{generic_cons}
  \nabla\cdot \underline{\mathsf{t}}(n)=\nabla\cdot(\underline{\mathsf{t}}_g(n)+\underline{\mathsf{t}}_m(n))=0,
\end{eqnarray}
where the formulae
\begin{eqnarray}\label{canonical_set}
  \begin{aligned}
    \underline{\mathsf{t}}_m(n)&=\partial_b\langle {\phi_i}_{,n}\partial_{{\phi_i}_{,b}}\mathproper{L}_m\rangle-n\mathproper{L}_m, \\
    \underline{\mathsf{t}}_g(n)&=\partial_b\langle \Omega(\partial_a)_{,n}\partial_{\Omega(a)_{,b}}\mathproper{L}_g\rangle-n\mathproper{L}_g,
\end{aligned}
\end{eqnarray}
are adapted from an early exposition on Lagrangian field theories using the spacetime algebra \cite{lasenby6}. In particular, the linear functions in \eqref{canonical_set} are the adjoints of those in \onlinecite{lasenby6} so as to agree with conventions regarding the stress energy tensor of the Dirac field in \onlinecite{doran}. Furthermore, following the conventions of \onlinecite{lasenby}, \textit{multivector} derivatives with respect to \textit{vector} derivatives of dynamical fields are difficult to work with, therefore the vector derivatives are `blunted' into directional derivatives by contracting with arbitrary constant basis vectors.
It should be emphasised that the term \textit{tensor} is used here in a loose sense. Whilst the notion of a tensor is perfectly well defined in the spacetime algebra, the quantities obtained from \eqref{canonical_set} are in no way constrained to be covariant. A non-covariant linear function may be interpreted as a pseudotensor.
\par
Now it is anticipated in \onlinecite{lasenby} and \onlinecite{doran} that in the case of Einstein-Hilbert gauge theory gravity, \eqref{generic_cons} does not yield any new information. If we explore this claim, we find that whilst the resultant conservation law is indeed a recycling of the field equations, it brings to light a curious class of identically conserved currents in the theory. For the gravitational sector, we can substitute \eqref{grav_sect} to give
\begin{eqnarray}
  \begin{aligned}
  &\kappa\underline{\mathsf{t}}_g(n)=\\
  &\phantom{==}-\partial_b\langle n\cdot\nabla\Omega(\partial_a)\partial_{\Omega(a)_{,b}}\langle\bar{\mathsf{h}}(\partial_d\wedge\partial_c)c\cdot\dot{\nabla}\dot{\Omega}(d)\rangle\rangle\det\underline{\mathsf{h}}^{-1}\\
  &\phantom{==}+\tfrac{1}{2}n\mathcal{R}\det\underline{\mathsf{h}}^{-1}\\
  &\phantom{=}=\partial_b\langle n\cdot\dot{\nabla}\dot{\Omega}(\partial_a)\bar{\mathsf{h}}(b\wedge a)\rangle\det\underline{\mathsf{h}}^{-1}+\tfrac{1}{2}n\mathcal{R}\det\underline{\mathsf{h}}^{-1}\\
  &\phantom{=}=\underline{\mathsf{h}}(\partial_a\cdot(n\cdot\dot{\nabla}\dot{\Omega}(\underline{\mathsf{h}}(a)))\det\underline{\mathsf{h}}^{-1}+\tfrac{1}{2}n\mathcal{R}\det\underline{\mathsf{h}}^{-1}.
  \end{aligned}
\end{eqnarray}
Now in the absence of torsion we can find the corresponding \textit{covariantly conserved} current using
\begin{eqnarray}
  \nabla\cdot J=0\implies \mathcal{D}\cdot \mathcal{J}=0,
\end{eqnarray}
where
\begin{eqnarray}\label{flat_current_formula}
  \mathcal{J}=\underline{\mathsf{h}}^{-1}(J)\det \underline{\mathsf{h}}.
\end{eqnarray}
For matter, it is easy to see that with a Lagrangian of the form \eqref{generic_mat}, the covariantised quantity is the functional stress-energy tensor\footnote{In the presence of spin, the $\Omega(a)$-dependence in the Lagrangian significantly complicates the picture.}
\begin{eqnarray}
  \underline{\mathsf{h}}^{-1}(\underline{\mathsf{t}}_m(n))\det\underline{\mathsf{h}}=\kappa\mathcal{T}(\underline{\mathsf{h}}^{-1}(n)),
\end{eqnarray}
whilst for gravity we have
\begin{equation}
  \begin{aligned}
    &\underline{\mathsf{h}}^{-1}(\underline{\mathsf{t}}_g(n))\det\underline{\mathsf{h}}=\\
    &\phantom{==}\partial_a\cdot(n\cdot\dot{\nabla}\dot{\Omega}(\underline{\mathsf{h}}(a))+\tfrac{1}{2}\underline{\mathsf{h}}^{-1}(n)\mathcal{R}\\
      &\phantom{=}=-\mathcal{R}(\underline{\mathsf{h}}^{-1}(n))+\tfrac{1}{2}\underline{\mathsf{h}}^{-1}(n)\mathcal{R}+\mathcal{D}\cdot\omega(\underline{\mathsf{h}}^{-1}(n)).
    \end{aligned}
  \label{}
\end{equation}
Assembling these results, we see that global spacetime translations give rise to the following conservation law
\begin{eqnarray}\label{conv_cons}
  \mathcal{D}\cdot[-\mathcal{G}(\underline{\mathsf{h}}^{-1}(n))+\kappa\mathcal{T}(\underline{\mathsf{h}}^{-1}(n))+\mathcal{D}\cdot\Omega(n)]=0.
\end{eqnarray}
As expected, this law does not tell us anything we did not already know. The first two terms in brackets can immediately be removed using the Einstein equation, \eqref{einstein_gtg}. Furthermore, the final term can also be removed using a law which applies to general bivectors, $B$, in the absence of torsion
\begin{eqnarray}\label{doubledot}
  \mathcal{D}\cdot(\mathcal{D}\cdot B)=0.
\end{eqnarray}
To prove this, let us write
\begin{eqnarray}
  \begin{aligned}
  &\mathcal{D}\cdot(\mathcal{D}\cdot B)=(\overrightarrow{\mathcal{D}}\wedge\mathcal{D})\cdot B\\
  &\phantom{=}=(\bar{\mathsf{h}}(\partial_a)\mathcal{D}_a\wedge\bar{\mathsf{h}}(\partial_b)\mathcal{D}_b)\cdot B\\
  &\phantom{=}=(\mathcal{D}\wedge\bar{\mathsf{h}}(\partial_b))\cdot\mathcal{D}_b B+\bar{\mathsf{h}}(\partial_a\wedge\partial_b)\cdot\overrightarrow{\mathcal{D}}_a\mathcal{D}_b B\\
  &\phantom{=}=(\partial_a\wedge\partial_b)\cdot(\mathcal{R}(a\wedge b)\times B),
\end{aligned}
\end{eqnarray}
where we have assumed $a$ to be an arbitrary constant. We can then apply some further identities to show that
\begin{eqnarray}
  \begin{aligned}
    &(\partial_a\wedge\partial_b)\cdot(\mathcal{R}(a\wedge b)\times B)=\\
    &\phantom{==}\mathcal{R}(a\wedge b)\cdot(B\times(\partial_a\wedge\partial_b))\\
    &\phantom{=}=\mathcal{R}(a\wedge b)\cdot( (B\cdot\partial_a)\wedge\partial_b-\partial_b\wedge(B\cdot\partial_b))\\
    &\phantom{=}=-2(B\cdot\partial_a)\cdot\mathcal{R}(a)=-2(\partial_a\wedge\mathcal{R}(a))\cdot B=0
\end{aligned}
\end{eqnarray}
and the last equality is a result of the symmetry of the Ricci tensor. Note that covariance of $B$ is not required. We can, in fact, arrive at \eqref{doubledot} through the powerful `double wedge' relation\footnote{See also Appendix \ref{a3}.} presented in \onlinecite{lasenby,doran} for arbitrary multivector, $M$, in the absence of torsion
\begin{eqnarray}
  \mathcal{D}\wedge(\mathcal{D}\wedge M)=0.
\end{eqnarray}
To see this we set $M=IB$ and use the pseudoscalar to convert between interior and exterior products
\begin{eqnarray}
  \mathcal{D}\wedge(\mathcal{D}\wedge IB)=-\overrightarrow{\mathcal{D}}\wedge I\mathcal{D}\cdot B=I\mathcal{D}\cdot(\mathcal{D}\cdot B).
\end{eqnarray}
Equation \eqref{doubledot} provides us with an instant formula for generating conserved vector currents in gauge theory gravity. Given covariant vectors  $\mathcal{U}$ and $\mathcal{V}$ we have
\begin{eqnarray}\label{recipe}
  \mathcal{J}=\mathcal{D}\cdot(\mathcal{U}\wedge\mathcal{V}).
\end{eqnarray}
In fact, only one vector field is necessary. Setting $\mathcal{U}=\mathcal{D}$ allows us to construct the conserved currents introduced by Komar in \onlinecite{komar}
\begin{eqnarray}\label{komar_current}
  \mathcal{J}=\mathcal{D}\cdot(\mathcal{D}\wedge\mathcal{V})-\dot{\mathcal{D}}\cdot(\dot{\mathcal{D}}\wedge\mathcal{V}).
\end{eqnarray}
It is clear from \eqref{komar_current} that from the gauge theory perspective, Komar currents are a composite of identically conserved currents and consequently, as a whole, are purely second order in the covariant derivatives of $\mathcal{V}$. This raises the question of whether currents corresponding to the first term in \eqref{komar_current} have any useful application, but we will not consider this further here.
Now the observations that have been made about the covariantised equation, \eqref{conv_cons}, could in principle be made just as well in the flat space. Indeed by writing $B$ as an exterior product of two arbitrary vectors (or a sum thereof), it is not hard to show that
\begin{eqnarray}
  \mathcal{D}\cdot B=\nabla\cdot(\underline{\mathsf{h}}(B)\det\underline{\mathsf{h}}^{-1}),
\end{eqnarray}
and in this way the term in question is revealed to be the identically conserved gradient of a \textit{superpotential} buried in the Einstein tensor
\begin{eqnarray}
  \nabla\cdot(\nabla\cdot\underline{\mathsf{h}}(\omega(\underline{\mathsf{h}}^{-1}(n)))\det\underline{\mathsf{h}}^{-1})=0.
\end{eqnarray}
As we mentioned in Section \ref{s1}, the identification of superpotentials which split the Einstein tensor has long been a fruitful approach to finding gravitational stress-energy tensors and pseudotensors. Thus we see that the variational approach to gravitational stress-energy localisation in Einstein-Hilbert gauge theory gravity introduces its own split in \eqref{conv_cons}. The gravitational stress-energy tensor (or pseudotensor) implied by \eqref{conv_cons} contains gradients of the rotational gauge field, which, once substituted for by \eqref{torsion_one} will become second derivatives of the displacement gauge field. In the next section we shall see how this may be avoided by an alternative choice of gravitational Lagrangian. 
\subsection{M{\o}ller's pseudotensor}\label{s3b}

The action \eqref{gtg_action} has the advantages of simplicity and covariance, but neither of these properties is necessary to reproduce the field equations. In obtaining his complex \eqref{einstein_comp}, Einstein removed any second derivatives of the metric appearing in the Einstein-Hilbert Lagrangian by means of a surface term, since a Lagrangian which is homogeneously second order in $\partial g$ must produce a canonical stress energy (affine) tensor with that same property. Similarly, one could use a surface term to lever the vector derivative off the rotation gauge fields and onto the displacement gauge fields in \eqref{gtg_action}, and reasonably hope that the resulting pseudotensor will have more desirable properties than that obtained above
\begin{eqnarray}
  \begin{aligned}
  \mathproper{L}_g(\bar{\mathsf{h}}(a),\Omega(a),\Omega(a)_{,b})=&_M\mathproper{L}_g(\bar{\mathsf{h}}(a),\bar{\mathsf{h}}(a)_{,b},\Omega(a))\\
  &+\nabla\cdot{_M\mathproper{F}}.
\end{aligned}
\end{eqnarray}
By inspection of \eqref{grav_sect}, the obvious `minimal' choice of surface term is simply
\begin{eqnarray}
  \kappa{_M\mathproper{F}}=\underline{\mathsf{h}}(\partial_a\cdot\omega(a))\det\underline{\mathsf{h}}^{-1},
\end{eqnarray}
with the new gravitational Lagrangian given by the formula
\begin{equation}\label{newlag}
  \begin{aligned}
    \kappa{_M\mathproper{L}_g}=&\left( \partial_a \wedge\partial_b\right) \cdot\big[\Omega(\underline{\mathsf{h}}(b))\nabla\cdot\left( \det \underline{\mathsf{h}}^{-1}\underline{\mathsf{h}}(a) \right)\\&+\det \underline{\mathsf{h}}^{-1} \underline{\mathsf{h}}(a)\cdot\dot{\nabla}\Omega(\dot{\underline{\mathsf{h}}}(b))\\&+\tfrac{1}{2}\Omega(\underline{\mathsf{h}}(a))\times\Omega(\underline{\mathsf{h}}(b))\det \underline{\mathsf{h}}^{-1}\big].
\end{aligned}
\end{equation}
The subscript reflects the fact that this most natural modification to the gravitational action has resulted in precisely the effective Lagrangian of M{\o}ller. This is not surprising, given that M{\o}ller was working at the level of the tetrad. Indeed, had we attempted to obtain Einstein's pseudotensor more carefully the necessary modifications to the gravitational Lagrangian might have appeared clumsy or unnatural. By applying \eqref{torsion_one} and \eqref{torsion_two} we find that \eqref{newlag} can be written in the very compact form identified by Hestenes in \onlinecite{hestenes}
\begin{eqnarray}\label{hestenes}
  \begin{alignedat}{2}
    \kappa{_M\mathproper{L}_g}&=&&(-(\dot{\bar{\mathsf{h}}}(\dot{\nabla})\cdot a-\bar{\mathsf{h}}(\dot{\nabla})\cdot a\partial_c\dot{\bar{\mathsf{h}}}\bar{\mathsf{h}}^{-1}(c))\partial_a\cdot(\partial_b\cdot\omega(b))\\
      & &&-(\bar{\mathsf{h}}(\dot{\nabla})\wedge\dot{\bar{\mathsf{h}}}\bar{\mathsf{h}}^{-1}(\partial_b))\cdot\omega(b)\\
    & &&+\tfrac{1}{2}(\partial_a\wedge\partial_b)\cdot(\omega(a)\times\omega(b)))\det\underline{\mathsf{h}}^{-1}\\
    &={}&&(  (\partial_a\cdot\omega(b))\cdot(\partial_b\cdot\omega(a))-(\partial_a\cdot\omega(a))^2\\
     & &&+\tfrac{1}{2}(\partial_a\wedge\partial_b)\cdot(\omega(a)\times\omega(b)))\det\underline{\mathsf{h}}^{-1}\\
     &={}&&-\tfrac{1}{2}(\partial_a\wedge\partial_b)\cdot(\omega(a)\times\omega(b))\det\underline{\mathsf{h}}^{-1}.
    \end{alignedat}
\end{eqnarray}
Since $_M\mathproper{L}_g\det\underline{\mathsf{h}}$ is dependent only on the $\omega(a)$-fields, we see that it is still position gauge covariant, having lost only the rotational gauge invariance of $\mathcal{R}$.
The symmetry of the new action under global spacetime translations again implies a conservation law on the flat background. This time it is convenient to evaluate the adjoint form corresponding to \eqref{canonical_set}
\begin{equation}\label{canonical_set_adj}
_M\bar{\mathsf{t}}_g(n)=\dot{\nabla}\langle\dot{\bar{\mathsf{h}}}(\partial_a)\partial_{\bar{\mathsf{h}}(a)_{,n}}{_M\mathproper{L}_g}\rangle -n{_M\mathproper{L}_g}.
\end{equation}
Though less coherent than \eqref{hestenes}, the expanded form \eqref{newlag} is far easier to work with, and the calculation requires only two steps. To find the contribution of the first term in \eqref{newlag} to that in \eqref{canonical_set_adj} we will need
\begin{equation}\label{first}
  \begin{aligned}
&   \dot{\nabla}\langle(\partial_a \wedge\partial_b)\cdot\Omega(\underline{\mathsf{h}}(b))\dot{\bar{\mathsf{h}}}(\partial_c)\partial_{\bar{\mathsf{h}}(c)_{,n}}( \det\underline{\mathsf{h}}^{-1}a\cdot\underline{\mathsf{h}}(\partial_d)_{,d}\\
&\phantom{==} +\partial_d\cdot\underline{\mathsf{h}}(a)(\det\underline{\mathsf{h}}^{-1})_{,d})\rangle\\
&\phantom{=}=\dot{\nabla}(\partial_a \wedge\partial_b)\cdot\omega(b)(a\cdot\dot{\bar{\mathsf{h}}}(n)\\
&\phantom{==}-n\cdot\underline{\mathsf{h}}(a)\partial_c\cdot\dot{\underline{\mathsf{h}}}\underline{\mathsf{h}}^{-1}(c))\det \underline{\mathsf{h}}^{-1},
  \end{aligned}
\end{equation}
where we make use of the identity from Appendix \ref{a4}
\begin{eqnarray}
  \partial_{\bar{\mathsf{h}}(c)_{,n}}(\det\underline{\mathsf{h}}^{-1})_{,b}=-(n\cdot b)\det\underline{\mathsf{h}}^{-1}\underline{\mathsf{h}}^{-1}(c). 
\end{eqnarray}
Meanwhile the second term in \eqref{newlag} contributes
\begin{equation}
  \begin{aligned}
  &\dot{\nabla}\bar{\mathsf{h}}(\partial_a)\cdot(\partial_b\cdot\Omega(\partial_d\dot{\bar{\mathsf{h}}}(\partial_c)\cdot(\partial_{\bar{\mathsf{h}}(c)_{,n}}b\cdot\bar{\mathsf{h}}(d)_{,a})))\det\underline{\mathsf{h}}^{-1}\\
  &\phantom{=}=\dot{\nabla}(\bar{\mathsf{h}}(n)\wedge\dot{\bar{\mathsf{h}}}\bar{\mathsf{h}}^{-1}(\partial_b))\cdot\omega(b)\det\underline{\mathsf{h}}^{-1}.
  \label{<+label+>}
\end{aligned}
\end{equation}
If we assemble these we arrive at the following formula for the pseudotensor of M{\o}ller expressed as a linear function in terms of the gravitational gauge fields
\begin{equation}\label{moller_explicit}
  \begin{aligned}
    &\kappa_M\bar{\mathsf{t}}_g(n)=\dot{\nabla}(\dot{\bar{\mathsf{h}}}(n)\wedge\partial_b-\bar{\mathsf{h}}(n)\wedge\partial_b \partial_c\cdot\dot{\bar{\mathsf{h}}}\bar{\mathsf{h}}^{-1}(c)\\
    &+\bar{\mathsf{h}}(n)\wedge\dot{\bar{\mathsf{h}}}\bar{\mathsf{h}}^{-1}(\partial_b))\cdot\omega(b)\det\underline{\mathsf{h}}^{-1}-\kappa n{_M\mathproper{L}_g}.
  \end{aligned}
\end{equation}
We will make use of this explicit formula in the final section. Note that the trace of the M{\o}ller pseudotensor reduces to his effective Lagrangian
\begin{eqnarray}
  \begin{aligned}
  \kappa\partial_n\cdot{_M\bar{\mathsf{t}}_g}&(n)\\
  &=3(\partial_a\wedge\partial_b)\cdot(\omega(a)\times\omega(b))\det\underline{\mathsf{h}}^{-1},
\end{aligned}
\end{eqnarray}
yet the pseudotensor itself cannot be expressed purely in terms of the $\omega(a)$: since M{\o}ller's superpotential is tensorial, the energy-momentum complex and pseudotensor cannot be. An alternative form for the pseudotensor is given in \onlinecite{hestenes}, and we compare the two in Appendix \ref{a2}.
As before, we will expect the conservation law on $M_4$ to be
\begin{eqnarray}
  _M\dot{\bar{\mathsf{t}}}(\dot{\nabla})=0, \quad {_M\bar{\mathsf{t}}}(n)={_M\bar{\mathsf{t}}_g}(n)+\bar{\mathsf{t}}_m(n),
\end{eqnarray}
but in order to obtain some very useful results we will arrive at this by the same circuitous route taken by Dirac when discussing the complex of Einstein in \onlinecite{Dirac}.
A very useful consequence of the field equations and the contracted Bianchi identity is the covariant conservation law \eqref{origin} with which we began our discussion
\begin{eqnarray}\label{bianchi_1}
  \dot{\mathcal{T}}(\dot{\mathcal{D}})=0.
\end{eqnarray}
This can be expanded as a vector derivative with two Levi-Civita connection terms
\begin{equation}
  \begin{alignedat}{2}
    \dot{\mathcal{T}}(\dot{\mathcal{D}})&={}&&\partial_a \dot{\mathcal{D}}\cdot\dot{\mathcal{T}}(a)\\
    &={}&&\partial_a \mathcal{D}\cdot\mathcal{T}(a)-\partial_a\partial_b\cdot\mathcal{T}(b\cdot\mathcal{D}a)\\
    &={}&&\partial_a\bar{\mathsf{h}}(\dot{\nabla})\cdot\dot{\mathcal{T}}(a)+\partial_a\partial_b\cdot[\omega(b)\times\mathcal{T}(a)]\\
    & &&-\partial_a\partial_b\cdot\mathcal{T}(b\cdot\partial_c \omega(c)\times a)\\
    &={}&&\dot{\mathcal{T}}(\bar{\mathsf{h}}(\dot{\nabla}))+\mathcal{T}(\partial_b\cdot\omega(b))-\mathcal{T}(\partial_c)\cdot\omega(c).
  \end{alignedat}
\end{equation}
Having performed all Palatini variations, we can eliminate the connection in terms of the displacement gauge field in the absence of torsion with \eqref{torsion_one}, and use the symmetry of the functional stress-energy tensor of matter, $\mathcal{T}(a)$, to write
\begin{eqnarray}
  \begin{aligned}
  \dot{\mathcal{T}}(\dot{\mathcal{D}})=&\dot{\mathcal{T}}(\bar{\mathsf{h}}(\dot{\nabla}))+\mathcal{T}(\dot{\bar{\mathsf{h}}}(\dot{\nabla}))\\
  &-\mathcal{T}(\bar{\mathsf{h}}(\dot{\nabla}))\partial_b\cdot\dot{\bar{\mathsf{h}}}\bar{\mathsf{h}}^{-1}(b)-\dot{\bar{\mathsf{h}}}\bar{\mathsf{h}}^{-1}\mathcal{T}(\bar{\mathsf{h}}(\dot{\nabla}))\\
  &+\bar{\mathsf{h}}(\dot{\nabla})\dot{\bar{\mathsf{h}}}\bar{\mathsf{h}}^{-1}(\partial_c)\cdot\mathcal{T}(c)=0.
\end{aligned}
\end{eqnarray}
Finally, by applying the displacement gauge field we can collect  some terms into a convenient \textit{total} divergence on $M_4$
\begin{eqnarray}
  \begin{aligned}
  \bar{\mathsf{h}}^{-1}(\dot{\mathcal{T}}(\dot{\mathcal{D}}))&\det\underline{\mathsf{h}}^{-1}=\bar{\mathsf{h}}^{-1}(\mathcal{T}(\bar{\mathsf{h}}(\overleftrightarrow{\nabla})))\det\underline{\mathsf{h}}^{-1}\\
  &+\dot{\nabla}\dot{\bar{\mathsf{h}}}\bar{\mathsf{h}}^{-1}(\partial_c)\cdot\mathcal{T}(c)\det\underline{\mathsf{h}}^{-1}=0.
\end{aligned}
\end{eqnarray}
So long as the matter is not a source of spin, the linear function acted on by this total divergence is seen to be its canonical stress-energy tensor, $\bar{\mathsf{t}}_m(a)=\bar{\mathsf{h}}^{-1}\mathcal{T}\bar{\mathsf{h}}(a)\det\underline{\mathsf{h}}^{-1}$.
The final term can then be equated with the exchange of energy-momentum with the gravitational field on $M_4$
\begin{equation}\label{moller_shortcut}
  \begin{alignedat}{2}
  {_M\dot{\bar{\mathsf{t}}}_g}(\dot{\nabla})&={}&&\dot{\nabla}\langle \partial_b\cdot\nabla(\dot{\bar{\mathsf{h}}}(\partial_a)\partial_{\bar{\mathsf{h}}(a)_{,b}}{_M\mathproper{L}_g})\rangle-\nabla{_M\mathproper{L}_g}\\
&={}&&\dot{\nabla}\langle\dot{\bar{\mathsf{h}}}(\partial_a)(\partial_b\cdot\nabla\partial_{\nabla\bar{\mathsf{h}}(a)_{,b}}{_M\mathproper{L}_g}\\
& &&-\partial_{\bar{\mathsf{h}}(a)}{_M\mathproper{L}_g})-\dot{\Omega}(\partial_a)\partial_{\Omega(a)}{_M\mathproper{L}_g} \rangle\\
    &={}&&\kappa\dot{\nabla}\dot{\bar{\mathsf{h}}}\bar{\mathsf{h}}^{-1}(\partial_a)\cdot\mathcal{T}(a)\det \underline{\mathsf{h}}^{-1},
\end{alignedat}
\end{equation}
where we have assumed for the final equality that, since the Lagrangian has been modified only by a surface term, the field equations should be unchanged
\begin{eqnarray}
  \begin{alignedat}{2}
      \mathcal{G}(\underline{\mathsf{h}}^{-1}(a))\det \underline{\mathsf{h}}^{-1}&={}&&\partial_b\cdot\nabla\partial_{\bar{\mathsf{h}}(a)_{,b}}{_M\mathproper{L}_g}\\
     & && -\partial_{\bar{\mathsf{h}}(a)}{_M\mathproper{L}_g},\\
    \kappa\mathcal{S}(\bar{\mathsf{h}}(a))\det \underline{\mathsf{h}}^{-1}&={}&&\partial_{\Omega(a)}{_M\mathproper{L}_g}.
\end{alignedat}
\end{eqnarray}
As a result we find that energy and momentum are conserved on $M_4$ in the expected manner
\begin{eqnarray}
  {_M\dot{\underline{\mathsf{t}}}_g}(\dot{\nabla})+\dot{\underline{\mathsf{t}}}_m(\dot{\nabla})=0,
\end{eqnarray}
but in the process we have equated the divergence of M{\o}ller's pseudotensor on $M_4$ to the final term in \eqref{moller_shortcut} -- this expression will shortly become very useful.
\subsection{The general Klein-Gordon correspondence}\label{s3c}

In Sections \ref{s2b} and \ref{s2c} we developed a natural way to extend the tensor of Butcher to the pseudotensor of Einstein. Now the correspondence between gauge theory gravity and Einstein-Cartan theory has led us to the pseudotensor of M{\o}ller rather than that of Einstein, but it is observed in \onlinecite{hestenes} that the two are equivalent in the spacetimes under discussion in this article. A first application of this result is that the Klein-Gordon correspondence of Section \ref{s2e}, displayed by Butcher's tensor in the Newtonian limit, fully survives the `nonlinearisation' to M{\o}ller's pseudotensor in the presence of strong gravitational fields. In this regime the matter density, $\rho$, is not expected to single-handedly generate the gravitational potential, $\varphi$: the whole of the matter stress-energy tensor acts as the source in any theory of gravity and so we anticipate that $P$ will play a part. Therefore let us denote the general source density for $\varphi$ by $\varrho$ such that\footnote{There is no such need to introduce a new symbol for $\varphi$.}
\begin{eqnarray}
  \varrho=\rho+\mathcal{O}(\lambda^2),
\end{eqnarray}
where $\lambda$ is the Newtonian parameter. We will soon see that this generalisation is insightful in the context of the relativistic mass functions discussed in Section \ref{s2e}. Now the first step is to understand what the Klein-Gordon theory on $\check{\mathcal{M}}$ discussed in Section \ref{s2d} looks like in the spacetime algebra on $M_4$. We stress that this is precisely the same field theory on Minkowski spacetime, but expressed using the apparatus of geometric algebra. The Lagrangian, \eqref{KG_lagrangian1}, will be
\begin{eqnarray}\label{klein_gordon_lagrangian_ga}
  _{KG}\mathproper{L}=(\nabla\varphi)^2-\kappa\varphi\varrho,
\end{eqnarray}
where we have used the symbol $_{KG}\mathproper{L}$ to reflect that this is is a Lagrangian density directly on $M_4$, and as such it is not necessarily gauge covariant. The canonical stress energy tensor, which is symmetric, can be partitioned into a field term and an interaction term
\begin{eqnarray}\label{klein_gordon_set_ga}
  \begin{aligned}
    _{KG}\mathsf{t}(n)&=2\nabla\varphi n\cdot\nabla\varphi-n(\nabla\varphi)^2+\kappa n\varphi\varrho\\
  &=\nabla\varphi n\nabla \varphi-\kappa n\varphi.
\end{aligned}
\end{eqnarray}
The equation of motion, \eqref{field_equation}, will then be simply
\begin{eqnarray}\label{klein_gordon_eom_ga}
  \nabla^2\varphi=-\kappa\varrho/2.
\end{eqnarray}
Rectangular isotropic coordinates corresponding to the line element \eqref{isotropic_line_element} are introduced through the following displacement gauge fields acting on an orthonormal basis
\begin{eqnarray}
  \underline{\mathsf{h}}^{-1}(\gamma_0)=e^{A/2}\gamma_0, \quad \underline{\mathsf{h}}^{-1}(\gamma_i)=e^{C/2}\gamma_i,
\end{eqnarray}
these are expected to go over to \eqref{weak_isotropic} in the Newtonian limit.
The choice of timelike Killing vector is then made for us
\begin{eqnarray}\label{killing_isotropic}
  \mathcal{K}=\underline{\mathsf{h}}^{-1}(\gamma_0)=g_0.
\end{eqnarray}
The use of isotropic coordinates enables us to make two simplifications. Firstly the displacement gauge field will be self-adjoint, so we can dispense with over/underbar notation. Secondly, because the spacetime is static, we can assume the action of all vector derivatives is purely spatial. In practice this is a considerable shortcut, for instance the connection is simply
\begin{eqnarray}
  \omega(b)=\Omega(\mathsf{h}(b))=\mathsf{h}(\dot{\nabla})\wedge\dot{\mathsf{h}}\mathsf{h}^{-1}(b).
\end{eqnarray}
Proceeding in this manner, M{\o}ller's pseudotensor given by the linear function in \eqref{moller_explicit} is seen to be symmetric, and can be expressed very quickly by direct calculation as
\begin{eqnarray}\label{moller_isotropic_1}
  \begin{aligned}
    &{_M\mathsf{t}_g}(n)=\tfrac{1}{4}e^{(A+C)/4}\big[  \big((\dot{A}+2\dot{C})(\ddot{A}+2\ddot{C})\\
    &-\dot{A}\ddot{A}-2\dot{C}\ddot{C}\big)\ddot{\nabla}\dot{\nabla}\cdot n-(\dot{C}\ddot{C}+2\dot{A}\ddot{C})\dot{\nabla}\cdot\ddot{\nabla} n \big].
  \end{aligned}
\end{eqnarray}
The picture is further simplified when we assume that the spacetime is equipped with symmetry such that 
\begin{eqnarray}
  \nabla A\wedge\nabla C=0,
\end{eqnarray}
and this is clearly the case for the spherically symmetric spacetimes under discussion. We then see that M{\o}ller's pseudotensor does indeed adopt the form of the field part of \eqref{klein_gordon_set_ga},
\begin{eqnarray}\label{moller_isotropic_2}
  {_M\mathsf{t}_g}(n)=\nabla\varphi n\nabla\varphi,
\end{eqnarray}
with the radial field strength associated with the gravitational scalar potential given by
\begin{eqnarray}\label{field_strength}
  \varphi'=\tfrac{1}{2}e^{(A+C)/4}\sqrt{C'^2+2A'C'}.
\end{eqnarray}
Since we are now using isotropic coordinates, the prime denotes differentiation with respect to $r$ in the line element \eqref{isotropic_line_element}, rather than the Schwarzschild radial coordinate $\bar{r}$ which was so useful in Section \ref{s2e}.
\par
We have uncovered a remarkably compact picture of gravitational energetics: the stress and energy of the gravitational field on $M_4$ coincides with that of a scalar field, $\varphi$. It should be stressed that as with the linearised version of this relationship, the link with the Lagrangian, \eqref{klein_gordon_lagrangian_ga} is completely formal: if we want to construct an equation of motion analogous to \eqref{klein_gordon_eom_ga}, we will have to assemble it by hand rather than from an Euler-Lagrange equation. This we will now do for the spherically symmetric perfect fluid. We start by introducing two gauge invariant quantities from the Einstein equations
\begin{eqnarray}
  \begin{aligned}
    \mathcal{X}&=g^0\cdot\mathcal{R}(g_0)=\tfrac{1}{2}\kappa(\rho+3P),\\
    \mathcal{Y}&=g^i\cdot\mathcal{R}(g_i)=-\tfrac{3}{2}\kappa(\rho-P).
\end{aligned}
\end{eqnarray}
Then by taking a linear combination we can form something similar to a Poisson equation for the gravitational potential
\begin{eqnarray}
  \begin{aligned}
    -\frac{4}{r}\left( 2\varphi'^2+r\varphi'\varphi'' \right)&=C'e^{(A+3C)/2}\left(3\mathcal{X}+\mathcal{Y}\right)\\
    &+A'e^{(A+3C)/2}\left(\mathcal{Y}-\mathcal{X}\right),
\end{aligned}
\end{eqnarray}
or taking advantage of the regularity condition of the spacetime,
\begin{eqnarray}\label{poisson_full}
  \nabla^2\varphi=\tfrac{\kappa}{8}e^{(A+3C)/2}\left[ \rho A'-3PC'\right]/\varphi'.
\end{eqnarray}
In fact this formula is not unique to the spherical case: it is the isotropic form of the general relation \eqref{moller_shortcut} which we went to lengths to obtain as part of the conservation law on $M_4$ for M{\o}ller's pseudotensor. An obvious consequence of \eqref{poisson_full} is that the $\varphi$ obeys the Laplace equation in a vacuum. Particularly, for the case of relativistic stars, it is easy to show that $\varphi$ in the Schwarzschild spacetime above the stellar surface appears to have been generated by the gravitational mass of the star
\begin{eqnarray}\label{newtonian_full}
  \varphi=-M_T/r.
\end{eqnarray}
Hence, we see that not only does the Klein-Gordon correspondence hold in the presence of strong gravitational fields, but the gravitational potential retains its Newtonian form!
We are now in a position to equate $\varrho$ with the RHS of \eqref{poisson_full}. Doing so, the familiar Newtonian formula \eqref{newtonian_full} then indicates that $\varrho$ describes a \textit{gravitational} mass density on $M_4$: this refinement could not be made in the Newtonian limit where $\varrho$ and $\rho$ were indistinguishable. We also see that our formula for $\varrho$ still makes reference to the displacement gauge fields and gravitational potential - of which it is the `source'. This could perhaps be interpreted as a reflection of gravitational self-coupling, but we will not take it too seriously, not least because our entire discussion still lacks gauge invariance. Finally, if we expand $\varphi$ in some Newtonian parameter $\lambda$, 
\begin{eqnarray}
  \varphi=\sum_{n=1}^\infty \varphi_n \lambda^n, \quad \varrho=\sum_{n=1}^\infty \varrho_n \lambda^n,
\end{eqnarray}
we can write the Newtonian limit of rectangular isotropic coordinates as
\begin{eqnarray}
  \begin{aligned}
    e^{A/2}&=1+2\varphi_1\lambda+\mathcal{O}(\lambda^2),\\
    e^{C/2}&=1-2\varphi_1\lambda+\mathcal{O}(\lambda^2),
\end{aligned}
\end{eqnarray}
where the Newtonian potential is simply
\begin{eqnarray}
  \nabla^2\varphi_1\lambda=4\pi\rho.
\end{eqnarray}
The Poisson-like equation then expands to give us
\begin{eqnarray}
  \begin{aligned}
  \varrho_1\lambda+\varrho_2\lambda^2&=\rho(1-2\varphi_1\lambda)+3P\\
  &=\rho e^{(A+3C)/2}+3P+\mathcal{O}(\lambda^3),
\end{aligned}
\end{eqnarray}
which is \textit{precisely} the local virial theorem we anticipated in \eqref{komar_virial}, only it is expressed in isotropic coordinates. To compare, a hypothetical localisation of gravitational mass, $\tilde{\varrho}$, designed to reproduce the \textit{conventional} virial theorem, \eqref{schwarzschild_proper_virial}, would instead obey
\begin{eqnarray}
  \begin{aligned}
  \tilde\varrho_1\lambda+\tilde\varrho_2\lambda^2&=\rho(1-3\varphi_1\lambda)-3P\\
  &=\rho e^{3C/2}-3P+\mathcal{O}(\lambda^3).
\end{aligned}
\end{eqnarray}
In this way we connect back to the definitions of relativistic mass discussed earlier. One might wonder on the basis of virial theorems if the particular gravitational mass density expressed by $\varrho$ is perhaps that of Komar. In fact the two are distinct, but we will see in the following section that they are related.
\subsection{Mass in gauge theory gravity}\label{s3d}

We will conclude our discussion of the gauge theory approach with an attempt to place the conserved mass $\mathscr{M}_T$ mentioned in \eqref{consone} and \eqref{flat_binding_energy} in the global picture. In Section \ref{s3b}, we introduced equation \eqref{bianchi_1} as a consequence of the contracted Bianchi identity and Einstein equations. For any vector field $V$ we have
\begin{eqnarray}
  \dot{\mathcal{D}}\cdot\dot{\mathcal{T}}(V)=0,
\end{eqnarray}
which is the compact gauge theory statement of the non-conservation of material energy-momentum currents discussed in Section \ref{s2a}. In particular, if the $V$ were taken to be any of the basis vectors $\gamma_\mu$ (or even a physically meaningful vector such as the four-velocity of a local observer, $v$), we can see that \eqref{noncons} is equivalent to the result
\begin{eqnarray}\label{noncons_two}
  \mathcal{D}\cdot\mathcal{T}(V)=\partial_a\cdot\mathcal{T}(a\cdot\mathcal{D}V)\neq 0.
\end{eqnarray}
The matter energy momentum currents \textit{are} conserved, however, if for $V$ we take any of those vectors $\mathcal{K}$ which embody the symmetries of a given spacetime through the Killing equation
\begin{eqnarray}\label{killing_vector}
  a\cdot(b\cdot\mathcal{D}\mathcal{K})=-b\cdot(a\cdot\mathcal{D}\mathcal{K}).
\end{eqnarray}
We can remove the offending term on the RHS of \eqref{noncons_two} by writing
\begin{equation}
	\begin{aligned}
	\partial_a \cdot \mathcal{T}(a\cdot\mathcal{D}\mathcal{K})&=\partial_a \cdot \mathcal{T}(\partial_b (b\cdot(a\cdot\mathcal{D}\mathcal{K})))\\
	&= -\partial_a \cdot \mathcal{T}(\partial_b (a\cdot(b\cdot\mathcal{D}\mathcal{K}))),
\end{aligned}
\end{equation}
which we then re-arrange using the linearity and symmetry of $\mathcal{T}(a)$
\begin{equation}
	\begin{aligned}
		\partial_a \cdot \mathcal{T}(a\cdot\mathcal{D}\mathcal{K})&= -(b\cdot\mathcal{D}\mathcal{K})\cdot\mathcal{T}(\partial_b)\\
	&= -\mathcal{T}(b\cdot\mathcal{D}\mathcal{K})\cdot\partial_b\\
	&= -\partial_a\cdot\mathcal{T}(a\cdot\mathcal{D}\mathcal{K}),
\end{aligned}
\end{equation}
leaving us with a fully covariant and covariantly conserved vector current
\begin{eqnarray}\label{current}
  \mathcal{D}\cdot \mathcal{T}(\mathcal{K})=0.
\end{eqnarray}
To further develop our picture on $M_4$, we can reverse the use of \eqref{flat_current_formula} in Section \ref{s3a} to construct a corresponding conserved current
\begin{eqnarray}\label{flat_current}
  \nabla\cdot\underline{\mathsf{h}}(\mathcal{T}(\mathcal{K}))\det\underline{\mathsf{h}}^{-1}=0.
\end{eqnarray}
We are used, in the spacetime algebra, to obtaining conserved charges from conservation laws. A timelike observer in $M_4$ has proper time, $\tau$, which can be used as a coordinate function to define timelike basis vectors $\mathsf{e}_\tau$ and $\mathsf{e}^\tau$. Since $\mathsf{e}_\tau$ is the four-velocity of the observer, it must be a unit vector. The conserved charge density associated with the current, \eqref{flat_current} over the whole spatial hypersurface, $\Sigma_\tau$, can be integrated
\begin{eqnarray}\label{to_look}
  \begin{aligned}
    \mathcal{Q}_T&=\int_{\Sigma_\tau} |\mathrm{d}^3x|\mathsf{e}^\tau\cdot\underline{\mathsf{h}}(\mathcal{T}(\mathcal{K}))\det\underline{\mathsf{h}}^{-1}\\
    &=\int_{\Sigma_\tau} |\mathrm{d}^3x|\mathcal{T}(\mathcal{K})\cdot g^\tau g_\tau\cdot(\underline{\mathsf{h}}^{-1}(\mathring{I})\cdot I^{-1})\\
    &=\int_{\Sigma_\tau}\langle \mathcal{P}_{\perp}(\mathcal{T}(\mathcal{K}))\underline{\mathsf{h}}^{-1}(\mathrm{d}^3x)I^{-1} \rangle
\end{aligned}
\end{eqnarray}
where $\mathcal{P}_{\perp}(a)$ is the gauge invariant rejection operator and $\mathring{I}$ the pseudoscalar associated with $\Sigma_\tau$ - both are defined in Appendix \ref{a1}. We have chosen a calligraphic script for $\mathcal{Q}_T$ because from \eqref{to_look} we can immediately write it in covariant form
\begin{eqnarray}\label{to_look2}
  \mathcal{Q}_T= \int_{\Sigma_t}\langle \mathcal{T}(\mathcal{K})\underline{\mathsf{h}}^{-1}(\mathrm{d}^3x) I^{-1} \rangle.
\end{eqnarray}
Now we have $\mathcal{Q}_T$ in gauge-covariant form it is easier to interpret. By applying Gauss' law to \eqref{to_look2} we see that $\mathcal{Q}_T$ is independent of our choice of $\Sigma_\tau$ because of \eqref{current}. This is entirely equivalent to our observation in $M_4$ that $\mathcal{Q}_T$ appears as a conserved charge in the theory. It is now obvious of course that we are dealing with the same quantity $Q_T$ mentioned in \eqref{consone}. The formula \eqref{to_look2} is covariant, but still depends on the normalisation of $\mathcal{K}$. The other relativistic mass which used $\mathcal{K}$ was that of Komar which we denoted by $\mathfrak{M}_T$ but found to be equal to the gravitational mass $M_T$ -- there we insisted on normalising the time-like Killing vector to unity at spatial infinity in order to take advantage of the Newtonian regime. The force applied by the observer at spatial infinity to suspend a unit mass with four-velocity $v=\mathcal{K}/|\mathcal{K}|$ is 
\begin{eqnarray}
  \mathcal{F}=v\cdot\mathcal{D}\mathcal{K}=v\cdot(\mathcal{D}\wedge\mathcal{K}),
\end{eqnarray}
where the Killing equation \eqref{killing_vector} can be used to recognise the presence of the bivector. In Section \ref{s2e} the Komar integral was performed in the hypersurface orthogonal to $\mathcal{K}$: we can set up a coordinate system that reflects this by taking
\begin{eqnarray}
  \mathcal{K}=\underline{\mathsf{h}}^{-1}{\mathsf{e}_t}=g_t,
\end{eqnarray}
and integrating in the surface $\Sigma_t$.
By applying Gauss' law, \eqref{to_use}, we can find the Komar mass within some bounded region $V$ in $\Sigma_t$
\begin{eqnarray}
  \begin{aligned}
    \mathfrak{M}=\oint_{\partial V} \langle & \mathcal{D}\wedge\mathcal{K} \underline{\mathsf{h}}^{-1}(\mathrm{d}^2x)I^{-1}   \rangle=\\
  &\int_V \langle \mathcal{D}\cdot(\mathcal{D}\wedge\mathcal{K})\underline{\mathsf{h}}^{-1}(\mathrm{d}^3x)I^{-1}  \rangle.
\end{aligned}
\end{eqnarray}
The next step is to replace the second covariant derivative of the Killing field using \eqref{ricci_killing}
\begin{eqnarray}
  \overrightarrow{\mathcal{D}}\cdot\mathcal{D}\mathcal{K}=\partial_a\mathcal{R}(a)\cdot\mathcal{K},
\end{eqnarray}
and then the total Komar mass is found by extending $V$ over the whole spatial hypersurface, $\Sigma_t$, giving
\begin{eqnarray}
  \begin{aligned}
    \mathfrak{M}_T=\int_{\Sigma_t} \langle \mathcal{R}(\mathcal{K})\underline{\mathsf{h}}^{-1}(\mathrm{d}^3x)I^{-1}  \rangle.
  \end{aligned}
\end{eqnarray}
If we adopt the same normalisation of $\mathcal{K}$ used above in our conserved charge, we recover the mass mentioned in Section \ref{s2e}
\begin{eqnarray}
  \mathcal{Q}_T=\mathscr{M}_T.
\end{eqnarray}
Now we claimed in Section \ref{s1} that a viable energy-momentum complex ought to account for $M_T$. We have also shown that $\mathscr{M}_T$ can be thought of as a conserved charge on $M_4$, and that the pseudotensor of M{\o}ller appears as the stress-energy tensor of a scalar field there. Postponing objections relating to the physicality of these results to Section \ref{s4}, we will therefore conclude by balancing the energy budget directly on $M_4$ as well, by re-introducing the isotropic coordinates and orthonormal basis vectors. This system of coordinates is of course an example of precisely the kind we have just been considering. The energy density of the gravitational field on the background is given by
\begin{eqnarray}
  _MU_g=\gamma_0\cdot{_M\mathsf{t}_g}(\gamma_0)=-{_M\mathproper{L}_g},
\end{eqnarray}
so from the definition \eqref{newlag} of M{\o}ller's effective Lagrangian,
\begin{eqnarray}
  \begin{aligned}
  \int_V|\mathrm{d}^3x|&{_MU_g}\\
  &=\int_V|\mathrm{d}^3x|(\tfrac{1}{2\kappa}\mathcal{R}\det\underline{\mathsf{h}}^{-1}-\nabla\cdot{_M\mathproper{F}}).
\end{aligned}
\end{eqnarray}
Now if we apply the Einstein equations to the Ricci scalar, we see
\begin{eqnarray}
  \mathcal{R}=\kappa(3P-\rho),
\end{eqnarray}
so it is possible to write
\begin{eqnarray}
  \begin{aligned}
    \int_V&|\mathrm{d}^3x|({_MU_g}+\rho\det\underline{\mathsf{h}}^{-1})=\\
    &\int_V|\mathrm{d}^3x|\tfrac{1}{2}(\rho+3P)\det\underline{\mathsf{h}}^{-1}-\oint_{\partial V}|\mathrm{d}^2x|\mathsf{e}_r\cdot{_M\mathproper{F}}.
\end{aligned}
\end{eqnarray}
The first term on the RHS is immediately identifiable as $M_T /2$ since it is equivalent to the Komar mass integral. For asymptotically flat systems, the second term on the RHS approaches $M_T/2$ when evaluated at spatial infinity, so we have
\begin{eqnarray}\label{energy_split}
  \int_{\Sigma_t}|\mathrm{d}^3x|({_MU_g}+\rho\det\underline{\mathsf{h}}^{-1})=M_T.
\end{eqnarray}
The second term on the LHS of \eqref{energy_split} will clearly integrate to $\mathscr{M}_T$, again by comparison with Section \ref{s2e}. It remains only to interpret M{\o}ller and Einstein's account of the energy sequestered in the gravitational field as the binding energy of the system, allowing us to justify \eqref{flat_binding_energy}, the energy relation
\begin{eqnarray}
  \mathscr{M}_B+\mathscr{M}_T=M_T.
\end{eqnarray}
\subsection{Example: Schwarzschild star}\label{s3e}

It is constructive to illustrate the picture of gravitostatic energetics we have been building using a simple system. Perhaps the simplest static spherically symmetric perfect fluid is that known as a \textit{Schwarzschild star}, which has a constant proper density $\rho=\rho_0$ and pressureless surface at $\bar{r}=\bar{R}$ in Schwarzschild-like coordinates. Below the stellar surface, the functions appearing in the Schwarzschild line element \eqref{schwarzschild_line_element} are
\begin{equation}
	\begin{aligned}
		e^{A/2}&=\frac{1}{2}\Big( 3\sqrt{1-2M_{\mathrm{T}}/\bar{R}}-\sqrt{1-2M/\bar{r}} \Big),\\
		e^{B/2}&=1/\sqrt{ 1-2M/\bar{r}}.
\end{aligned}
\end{equation}
In terms of the Newtonian parameter $\lambda^{-1}=\bar{R}/M_{\mathrm{T}}$, the star has proper mass
\begin{equation}
  \begin{aligned}\label{schwarzschild_proper}
  	\mathproper{M}_{T}=&\frac{3}{8}\sqrt{\lambda^{-1}}M_{\mathrm{T}}\Big[-2\sqrt{\lambda^{-1}-2}\\&+\sqrt{2}\lambda^{-1}\tan^{-1}\left(\sqrt{2/\left( \lambda^{-1}-2 \right)} \right) \Big],
      \end{aligned}
\end{equation}
and conserved mass
\begin{equation}\label{schwarzschild_gtg}
\begin{aligned}
  &\mathscr{M}_{T}=\frac{1}{16}M_{\mathrm{T}}\Big[-18\lambda^{-1}+28\\&+9\sqrt{2}\lambda^{-1}\sqrt{\lambda^{-1}-2}\tan^{-1}\left(\sqrt{2/\left( \lambda^{-1}-2 \right)} \right) \Big].
\end{aligned}
\end{equation}
In these formulae, the Schwarzschild coordinate mass function is simply $M=M_{\mathrm{T}}\bar{r}^3/\bar{R}^3$. Superficially the functions \eqref{schwarzschild_proper} and \eqref{schwarzschild_gtg} appear very similar, and indeed both agree on the Newtonian limit of the binding energy
\begin{eqnarray}
  \mathproper{M}_{B}=\mathscr{M}_{B}=3M_{T}/5\lambda^{-1}+\mathcal{O}(\lambda^{2}).
\end{eqnarray}
As is shown in Figure \ref{figure-2} however, $\mathproper{M}_{T}>M_{T}>\mathscr{M}_{T}$, so we see that in the alternative interpretation, \eqref{flat_binding_energy}, the gravitational mass loses a positive binding energy.
\par
To see the Klein-Gordon correspondence in action, we must base our displacement gauge fields beneath the stellar surface on the isotropic coordinates first set out by Wyman \cite{wyman} in 1946. The functions appearing in the line element \eqref{isotropic_line_element} are
\begin{equation}\label{sint}
\begin{aligned}
e^{A/2}&=\frac{2R-2M_{\mathrm{T}}+M_{\mathrm{T}}(4R-M_{\mathrm{T}})r^2/2R^3}{(2R+M_{\mathrm{T}})(1+M_{\mathrm{T}}r^2/2R^3)},\\
e^{C/2}&=\frac{(1+M_{\mathrm{T}}/2R)^3}{1+M_{\mathrm{T}}r^2/2R^3}.
\end{aligned}
\end{equation}
By substituting \eqref{sint} into \eqref{field_strength} we find the radial `gravitational field strength' beneath the stellar surface to be
\begin{equation}\label{fieldstrength}
	\begin{aligned}
	\varphi'&=\\
	&\frac{M_{\mathrm{T}}\left(1+\frac{M_{\mathrm{T}}}{2R}\right)r\sqrt{1+\frac{2M_{\mathrm{T}}}{R}-\frac{M_{\mathrm{T}}r^2}{R^3}\left(1-\frac{M_{\mathrm{T}}}{4R}\right)}}{R^3\left(1+M_{\mathrm{T}}r^2/2R^3\right)^2}.
\end{aligned}
\end{equation}
Figure \ref{figure-1} shows a pair of Schwarzschild stars with the same gravitational mass, $M_T$, but which have stalled their collapse at different isotropic radii. The integrated gravitational potential, $\varphi$, takes the same Newtonian form, \eqref{newtonian_full}, above the surface of each star. The source density, $\varrho$, does not share the uniform distribution of the proper mass: we see that the localisation of gravitational matter it represents tends to accumulate at the stellar core.
\section{Conclusions}\label{s4}

In the opening sections we used $\tau_{ab}$ as something of a springboard for the discussion of gravitational energy-momentum pseudotensors. Nonetheless, Sections \ref{s2a}, \ref{s2b} and \ref{s2c} would still benefit from some summary remarks. 
\par
It is difficult to gather from \onlinecite{butcher1,butcher2,butcher3,butcher4} a single \textit{motivating} definition of $\tau_{ab}$ which invites generalisation to nonlinear gravity. In many ways, the stress-energy and spin tensors emerge as simultaneously satisfying a long series of physically motivated requirements, any of which could be the focus of a generalisation attempt. Among these are the symmetry of $\tau_{ab}$ and corresponding conservation of angular momentum, gauge invariance (albeit restricted to plane gravitational waves) and satisfaction of the weak and dominant energy conditions. In this article we have explored only one such avenue: the total conservation of energy-momentum between matter and gravity in metrical general relativity.
As we have emphasised already, the \textit{metric} is no longer considered the fundamental dynamical variable of gravity, and it may be that our attempt has been akin to expanding a function $f(x)$ of no particular parity in $x^2$. 
We note that the use of tetrads in the nonlinearisation procedure is even suggested in \onlinecite{butcher4}, which we have barely considered in this article. There are perhaps two points to take away from our analysis
\begin{enumerate}
  \item \label{1} In the harmonic gauge, the linearised pseudotensor of Einstein is equivalent to $\tau_{ab}$ up to an identically conserved gauge current.
  \item \label{2} The original conservation law obeyed by $\tau_{ab}$ does not admit a symmetric, third-order correction to $\tau_{ab}$, quadratic in the first derivatives of the metric perturbation under the suggested perturbation schemes.
\end{enumerate}
Given Point \ref{2}, the failure at fourth order is irrelevant, since the required conservation law no longer holds. It is only by relaxing the conditions on the form of the conservation law to include an affine connection that we are able to make progress, and in doing so the missing component converges very rapidly order-by-order on the Christoffel symbols. This ultimately brings us to Point \ref{1} and the pseudotensor of Einstein, the uniqueness of which under various conditions has been established independently many times. The suggestion that the incompatibility of Einstein's pseudotensor with the equivalence principle could be tamed by the harmonic coordinate condition was made by M{\o}ller (in reference \cite{moller1} to the work of Fock) as early as 1961. In the linear regime, we may conclude that his remarks have proved insightful.
\vspace{5mm}
\par
It is now apparent, further to the work of \onlinecite{butcher2}, that the Klein-Gordon correspondence is a strong-field phenomenon in certain symmetric spacetimes, and applies to at least three formalisms for localising gravitational stress and energy. 
The strong-field extension is particularly interesting: the energetics of the gravitational field naturally identify a scalar field $\varphi$ as the \textit{gravitational potential}, which retains its simple Newtonian form above the surface of the densest Neutron star (provided such a star is not spinning).
It has been shown (see for example the review of Xulu \cite{xulu} and the references therein) that several energy-momentum complexes are in agreement in a wide class of spacetimes under quasi-Cartesian coordinates -- suggesting perhaps that the correspondence has a broader demographic than we have considered. Furthermore, the restriction to static spacetimes may prove artificial: isotropic coordinates, known as \textit{Weyl's canonical coordinates}, are very useful in describing stationary axisymmetric spacetimes \cite{stephani}. Conceivably, a generalisation of $\varphi$ to stationary spacetimes might be reminiscent of gravitoelectromagnetism. A necessary part of the picture appears to be the flat background provided by the gauge theory. To illustrate this, we note that the components of the M{\o}ller pseudotensor as evaluated in Einstein-Cartan theory using isotropic coordinates, $\{y^\mu\}$, can be recovered in a standard way by means of gravity frames

\begin{eqnarray}
  _Mt_{\mu\nu}=\nabla\varphi g_\mu \nabla\varphi\cdot g_\nu \det \underline{\mathsf{h}}.
\end{eqnarray}
In this form they do not clearly correspond to the field portion of the Klein-Gordon stress energy tensor. This also demonstrates the disadvantage of working entirely with gravity frames: if every available basis vector in an expression has been wrapped in a displacement gauge field for the sake of maintaining covariance, the construction on $M_4$ illustrated in Figure \ref{figure-1} may become somewhat obscured. Alternatively, to see the \textit{advantage} of gravity frames, one need only compare the form of M{\o}ller's pseudotensor obtained from Hestenes' unitary form in \onlinecite{hestenes} with our frame-free translation provided in the Appendix, \eqref{whole}. The latter has lost all of the elegance of the former and is difficult to comprehend. Perhaps more natural to the frame-free picture is the variational approach to localising gravitational energetics. Whilst the canonical pseudotensor of Einstein-Hilbert gauge theory gravity probably has undesirable qualities (and almost certainly already has a counterpart in classical\footnote{By \textit{classical} we mean the geometric picture set in space with curvature and torsion.} Einstein-Cartan theory), the derivation in Section \ref{s3a} has provided us with a recipe, \eqref{recipe}, for constructing conserved currents in gauge theories with geometric algebra. The variational approach has clear advantages over the use of tetrads in the derivation of M{\o}ller's pseudotensor: the problematic action of $\nabla$ on the rotational gauge fields in $\mathproper{L}_g$ is made manifest by the notation\footnote{In fact the necessary steps seemed to agree so entirely with the spirit of Einstein's approach in the metrical theory that one of us initially mistook $_M\mathproper{L}_g$ to be $_E\mathproper{L}_g$ until coming into contact with \onlinecite{hestenes}.}. Furthermore, we note that the use of the $\omega(a)$-fields is very convenient for verifying the tensorial nature of quantities.
\par
As an undercurrent to this discussion we have remarked on the relationship reflected in Figure \ref{figure-2}, that the conserved mass $\mathscr{M}_T$ and the proper mass $\mathproper{M}_T$ of relativistic stars appear to somehow `mirror' each other across the gravitational mass, $M_T$. Specifically we have observed 
\begin{enumerate}[label=\Roman*,ref=\Roman*]
  \item \label{o1} The binding energies $\mathscr{M}_B$ and $\mathproper{M}_B$ correspond in the Newtonian limit.
  \item \label{o2} The factor of $\rho+3P$ in the Komar density is suggestive of a \textit{local} virial theorem satisfied by $\mathscr{M}_T$ rather than $\mathproper{M}_T$ -- the later satisfies a \textit{global} virial theorem.
  \item \label{o3} On the $M_4$ background of gauge theory gravity, certain gravitational stress-energy pseudotensors in certain spacetimes under isotropic coordinates imitate the stress-energy tensor of a scalar field, $\varphi$, which appears to be generated by a gravitational mass density $\varrho$. This density replicates the same local virial theorem as the Komar density.
  \item \label{o4} The energy budget of the same pseudotensors on $M_4$ takes the form
    \begin{eqnarray}
    \mathscr{M}_T+\mathscr{M}_B=M_T.
  \end{eqnarray}
\end{enumerate}
Aside from the opposing sense in which the binding energy is `lost' in either picture, it is worth noting that $\mathproper{M}_B$ and $\mathscr{M}_B$ can also differ significantly in magnitude, though it is not at all apparent from Figure \ref{figure-2}. For example, the binding energy of the Earth, which well approximates a Schwarzschild star, is revised either way by $\num{1.2d8}\si{\gram}$ -- nearly the mass of a blue whale.
Fortunately, Points \ref{o1}, \ref{o2} and \ref{o4} are very far from subtle observations\footnote{Indeed, Point \ref{o1} is evident from Schwarzschild's 1916 solutions to the field equations, furthermore six decades have passed since Komar introduced his mass integral, followed shortly by M{\o}ller's proposal for an energy-momentum complex based on tetrads. For a dedicated treatment of the gravitational virial theorem, see e.g. \cite{jiri_4}.}, and doubtless the alternative picture of binding energy has been thoroughly interpreted elsewhere. Our main contribution is therefore Point \ref{o3} and the Klein-Gordon correspondence, though it is perhaps the least complete: the $M_4$ background is not observable, because all observers are bound by the gauge fields. Equivalently, gauge theory gravity is \textit{not} an {\ae}ther theory. Furthermore, the Klein-Gordon correspondence presently relies on a privileged isotropic coordinate system and we make no attempt in this article to construct a covariant generalisation (such an objection applies to stress-energy pseudotensors in general). With this in mind, and pending such an investigation, it remains to be seen whether the quantities $\varphi$ and $\varrho$ may find some physical significance beyond the mathematical structure of gauge theory gravity that suggests them.

\begin{acknowledgments}
  We are grateful to DO Hestenes for connecting us with his previous treatment of this topic \cite{hestenes}. WEVB is supported by STFC.  \end{acknowledgments}

\appendix
\section{Gauge invariance for Gauss' law}\label{a1}
In the classical formulation of general relativity, `gravitation' is equivalent to the geometry of the Riemann space, $V_4$, in which the physics unfolds. This geometry is intimately connected to the general integral theorem known as Gauss' (or Stokes') theorem, which is expressed using differential forms. Consequently, this theorem is often used to invoke gravitational effects when obtaining physical laws, with the derivation of the Komar mass in Section \ref{s2e} being a prime example. If the same laws arise in gauge theories of gravity we may wonder how the comparatively `informationless' integral theorem on flat Minkowski space, $M_4$, can be of any use in obtaining them. This paradox is resolved by insisting on the gauge invariance of all directed integrals. The first example of this principle is the factor of $\det\underline{\mathsf{h}}^{-1}$ in the action \eqref{gtg_action} which is an integral over the whole of $M_4$: this factor has an entirely non-trivial effect on the equations of motion.
\par
The next-simplest case is that of a covariant vector $\mathcal{J}$ integrated over an $n=3$ hypersurface $\partial V$ of directed measure $\mathrm{d}^3 x$ enclosing the $n=4$ volume $V$. We can think of the integrand as a linear function, $\mathsf{L}(a)$, of that measure
\begin{eqnarray}\label{surface_one}
  \oint_{\partial V}\mathsf{L}(\mathrm{d}^3x)=\oint_{\partial V}\langle \mathcal{J} \underline{\mathsf{h}}^{-1}(\mathrm{d}^3 x) I^{-1} \rangle.
\end{eqnarray}
The action of the displacement gauge field on the directed measure in \eqref{surface_one} guarantees gauge invariance of the hypersurface integral. Next we apply the fundamental theorem of geometric calculus in $M_4$ to obtain an integral with directed measure $\mathrm{d}^4 x$ on $V$, and expand in the following manner
\begin{eqnarray}\label{longer}
  \begin{aligned}
  &\oint_{\partial V}\mathsf{L}(\mathrm{d}^3 x)=\int_{V}\dot{\mathsf{L}}(\dot{\nabla}\mathrm{d}^4 x)\\
  &\phantom{=}=\int_{V}\langle \mathcal{J} \underline{\mathsf{h}}^{-1}(\overleftrightarrow{\nabla}\mathrm{d}^4 x) I^{-1} \rangle\\
  &\phantom{=}=\int_{V}\langle \mathcal{J}I \underline{\mathsf{h}}^{-1}(I^{-1}\overleftrightarrow{\nabla})|\mathrm{d}^4 x| \rangle\\
  &\phantom{=}=\int_{V}|\mathrm{d}^4 x| \mathcal{J}\cdot\bar{\mathsf{h}}(\overleftrightarrow{\nabla}) \det\underline{\mathsf{h}}^{-1}
  =\int_{V}|\mathrm{d}^4 x| \mathcal{D}\cdot\mathcal{J} \det\underline{\mathsf{h}}^{-1},
\end{aligned}
\end{eqnarray}
where the final equality follows from \eqref{torsion_two}.
Thus we see how the divergence in the volume naturally inherits the covariant derivative (and by extension, rotational gauge field) from the gauge invariance of the directed measure on the surface. 
\par
Of course, the integral theorem of differential geometry is not confined to hypersurfaces. In particular, we want to consider directed integrals \textit{within} a Cauchy hypersurface $\Sigma_t$ whose arbitrary geometry depends on its embedding in $M_4$. Fortunately the fundamental theorem of geometric calculus has been generalised to \textit{precisely} this contingency in the \textit{vector manifold} theory of Hestenes and Sobczyk. 
In this paper we are considering static spacetimes, it is possible to choose a $\Sigma_t$ everywhere orthogonal to the Killing field $\mathcal{K}$. 
To maintain generality, we will instead set up coordinates $x^\mu$ such that $x^0=t$ and an orthonormal basis is given by
\begin{eqnarray}
  \begin{aligned}
  &\mathsf{e}_\mu=\partial_\mu x, \quad \mathsf{e}^\nu=\nabla x^\nu,\\
  &\mathsf{e}_\mu\cdot\mathsf{e}^\nu=\delta^\nu_\mu, \quad \mathsf{e}_\mu\cdot\mathsf{e}_\nu=\eta_{\mu\nu}, \quad \mathsf{e}^\mu\cdot\mathsf{e}^\nu=\eta^{\mu\nu}.
  \end{aligned}
\end{eqnarray}
The pseudoscalar on $\Sigma_t$ can be defined
\begin{eqnarray}\label{proj_pse}
  \mathring{I}=\mathsf{e}^t\cdot I,
\end{eqnarray}
and following the convention of \onlinecite{doran} we will write the projection of the vector derivative onto $\Sigma_t$ as
\begin{eqnarray}\label{proj_der}
  \mathring{\nabla}=\partial.
\end{eqnarray}
The quantities in \eqref{proj_pse} and \eqref{proj_der} are not gauge covariant, but we will only invoke them as an intermediate step to equate gauge covariant quantities. 
Since $\mathsf{e}^t$ belongs to the cotangent space on $M_4$, the vector
\begin{eqnarray}
  g^t=\bar{\mathsf{h}}(\mathsf{e}^t)
\end{eqnarray}
is covariant, and always orthogonal to the covariantised tangent vectors in $\Sigma_t$,
\begin{eqnarray}
  g_i=\underline{\mathsf{h}}^{-1}(\mathsf{e}_i).
\end{eqnarray}
In the example in Section \ref{s3d}, we are actually concerned with the integral of a covariant bivector $\mathcal{B}$ over a closed hypersurface $\partial V$ within $\Sigma_t$ which surrounds some star
\begin{eqnarray}
  \oint_{\partial V}\mathsf{M}(\mathrm{d}^2x)=\oint_{\partial V}\langle\mathcal{B}\underline{\mathsf{h}}^{-1}(\mathrm{d}^2 x)I^{-1} \rangle.
\end{eqnarray}
If we apply the fundamental theorem of geometric calculus to the embedded integral we have
\begin{eqnarray}\label{long}
  \begin{aligned}
    &\oint_{\partial V}\mathsf{M}(\mathrm{d}^2x)=\int_V\dot{\mathsf{M}}(\dot{\partial}\mathrm{d}^3x)\\
    &\phantom{=}=\int_V\langle\mathcal{B}\underline{\mathsf{h}}^{-1}(  (\overleftarrow{\nabla}\cdot\mathring{I}\mathring{I}^{-1})\mathring{I} )I^{-1}|\mathrm{d}^3x|\rangle\\
    &\phantom{=}=\int_V \langle \mathcal{B}I\underline{\mathsf{h}}^{-1}(I^{-1}(\overleftrightarrow{\nabla}\wedge \mathsf{e}^t))|\mathrm{d}^3 x|   \rangle\\
    &\phantom{=}=\int_V |\mathrm{d}^3x|\bar{\mathsf{h}}(\overleftrightarrow{\nabla})\cdot(\bar{\mathsf{h}}(\mathsf{e}^t)\cdot\mathcal{B})\det\underline{\mathsf{h}}^{-1}\\
    &\phantom{=}=\int_V|\mathrm{d}^3x|\mathcal{D}\cdot(g^t\cdot\mathcal{B})\det\underline{\mathsf{h}}^{-1},
  \end{aligned}
\end{eqnarray}
where for the second equality we used
\begin{eqnarray}
  \nabla\wedge\mathsf{e}^t=0.
\end{eqnarray}
We can also think of \eqref{long} as an integral of the gauge-covariant rejection of $\mathcal{D}\cdot\mathcal{B}$ off $\Sigma_t$ over the bounded region
\begin{eqnarray}\label{to_use}
  \begin{aligned}
  \oint_{\partial V}\langle&\mathcal{B}\underline{\mathsf{h}}^{-1}(\mathrm{d}^2 x)I^{-1} \rangle\\
  &=\int_V (\mathcal{D}\cdot \mathcal{B})\cdot g^t g_t\cdot(\underline{\mathsf{h}}^{-1}(\mathrm{d}^3x)I^{-1})\\
 &=\int_V \langle \mathcal{P}_\perp (\mathcal{D}\cdot \mathcal{B})\underline{\mathsf{h}}^{-1}(\mathrm{d}^3x)I^{-1} \rangle.
\end{aligned}
\end{eqnarray}
A final simplification can be made by noticing that the gravity frames form a complete set, and so the identity operator may be defined as
\begin{eqnarray}\label{identity_operator}
  g^\mu \cdot(\mathcal{D}\cdot\mathcal{B}) g_\mu=\mathcal{D}\cdot\mathcal{B},
\end{eqnarray}
it is then easy to see that $\mathcal{P}_\perp (\mathcal{D}\cdot\mathcal{B})$ is the only term in \eqref{identity_operator} that survives inside the scalar part of the integrand in \eqref{to_use}, so we have
\begin{eqnarray}\label{to_use}
  \begin{aligned}
  \oint_{\partial V}\langle&\mathcal{B}\underline{\mathsf{h}}^{-1}(\mathrm{d}^2 x)I^{-1} \rangle=\int_V \langle \mathcal{D}\cdot \mathcal{B}\underline{\mathsf{h}}^{-1}(\mathrm{d}^3x)I^{-1} \rangle.
\end{aligned}
\end{eqnarray}
Note that \eqref{longer} may also be written in this `scalar part' form.
\section{Unitary form of M{\o}ller's pseudotensor}\label{a2}

We would like to verify that the linear machine in \eqref{moller_explicit} derived through variational principles is indeed the same as that arrived at in \onlinecite{hestenes} using the unitary form. In our frame-free notation, this form may be written as a sum of two terms, with the second term containing a single displacement gauge field gradient
\begin{eqnarray}\label{whole}
  \begin{aligned}
    &\kappa{_M\bar{\mathsf{t}}_g}(n)\det\underline{\mathsf{h}}\\
    &=\partial_c\bar{\mathsf{h}}(n\wedge\partial_a\wedge\partial_b)\cdot[\tfrac{1}{2}(\Omega(a)\times\Omega(b))\wedge\underline{\mathsf{h}}^{-1}(c)\\
    &\phantom{=}+\Omega(b)\wedge(\Omega(a)\cdot\underline{\mathsf{h}}^{-1}(c))]\\
    &\phantom{=}+\partial_c\bar{\mathsf{h}}(n\wedge\partial_a\wedge\partial_b)\cdot(\Omega(b)\wedge\underline{\mathsf{h}}^{-1}(c)_{,a}). 
  \end{aligned}
\end{eqnarray}
To prove the equivalence, we start by expanding this second term
\begin{eqnarray}\label{kinetic_part}
  \begin{aligned}
    &\partial_c\bar{\mathsf{h}}(n\wedge\partial_a\wedge\partial_b)\cdot(\Omega(b)\wedge\underline{\mathsf{h}}^{-1}(c)_{,a})\\
    &\phantom{=}=\dot{\bar{\mathsf{h}}}^{-1}(\omega(b)\cdot(\bar{\mathsf{h}}(\dot{\nabla})\wedge\partial_b\wedge\bar{\mathsf{h}}(n)))\\
    &\phantom{=}=\bar{\mathsf{h}}^{-1}\dot{\bar{\mathsf{h}}}(\dot{\nabla})(\bar{\mathsf{h}}(n)\wedge\partial_b)\cdot\omega(b)\\
    &\phantom{= =}-\dot{\nabla}(\dot{\bar{\mathsf{h}}}(n)\wedge\partial_b)\cdot\omega(b)\\
    &\phantom{= =}-\dot{\nabla}(\bar{\mathsf{h}}(n)\wedge\dot{\bar{\mathsf{h}}}\bar{\mathsf{h}}^{-1}(\partial_b))\cdot\omega(b)\\
    &\phantom{= =}+\bar{\mathsf{h}}^{-1}[(\bar{\mathsf{h}}(\dot{\nabla})\wedge\dot{\bar{\mathsf{h}}}(n))\cdot(\partial_b\cdot\omega(b))\\
    &\phantom{= =}-(\bar{\mathsf{h}}(\dot{\nabla})\wedge\dot{\bar{\mathsf{h}}}\bar{\mathsf{h}}^{-1}(\partial_b))\cdot(\bar{\mathsf{h}}(n)\cdot\omega(b))].
  \end{aligned}
\end{eqnarray}
By substituting for various displacement gauge field gradients using \eqref{torsion_one} and \eqref{torsion_two} we find
\begin{eqnarray}\label{other}
  \begin{aligned}
    &\partial_c\bar{\mathsf{h}}(n\wedge\partial_a\wedge\partial_b)\cdot(\Omega(b)\wedge\underline{\mathsf{h}}^{-1}(c)_{,a})\\
  &\phantom{=}=\kappa({_M\bar{\mathsf{t}}_g}(n)+n{_M\mathproper{L}_g})\det\underline{\mathsf{h}}\\
    &\phantom{= =}+\bar{\mathsf{h}}^{-1}[(\bar{\mathsf{h}}(n)\cdot\omega(b))\cdot(\partial_c\wedge(\partial_b\cdot\omega(b)))\\
      &\phantom{= =}-(\partial_b\cdot\omega(b))\cdot(\partial_c\wedge(\bar{\mathsf{h}}(n)\cdot\omega(b)))\\
    &\phantom{= =}+(\bar{\mathsf{h}}(n)\wedge\partial_b)\cdot\omega(b)(\partial_c\cdot\omega(c))],
  \end{aligned}
\end{eqnarray}
thus the second term in \eqref{whole} neatly collects the three `kinetic' terms in \eqref{moller_explicit}, leaving a remainder which may be expressed purely in terms of the $\omega(a)$-fields. If we now turn to the first term in \eqref{whole}, which is identified as \textit{M{\o}ller's superpotential}, we have
\begin{eqnarray}\label{other2}
  \begin{aligned}
    &\partial_c\bar{\mathsf{h}}(n\wedge\partial_a\wedge\partial_b)\cdot[\tfrac{1}{2}(\Omega(a)\times\Omega(b))\wedge\underline{\mathsf{h}}^{-1}(c)\\
    &\phantom{=}+\Omega(b)\wedge(\Omega(a)\cdot\underline{\mathsf{h}}^{-1}(c))]\\
  &=-\kappa n{_M\mathproper{L}_g}\det\underline{\mathsf{h}}\\
    &\phantom{=}-\tfrac{1}{2}(\bar{\mathsf{h}}(n)\cdot(\omega(a)\times\omega(b)))\cdot(\partial_a\wedge\partial_b)\\
    &\phantom{=}-\omega(a)\cdot[\omega(b)\cdot(\bar{\mathsf{h}}(n)\wedge\partial_a\wedge\partial_b)].
  \end{aligned}
\end{eqnarray}
After expanding, the trailing terms in \eqref{other} and \eqref{other2} cancel exactly.
\section{Killing fields and the first Bianchi identity}\label{a3}
In Section \ref{s3a} we mention the `double wedge' equation
\begin{eqnarray}\label{double_wedge}
  \mathcal{D}\wedge(\mathcal{D}\wedge M)=0,
\end{eqnarray}
which is true for \textit{any} multivector $M$. By contracting with an arbitrary constant basis trivector and expanding the resultant scalar, it is easy to see that \eqref{double_wedge} is a statement of the first Bianchi identity, and hence a symmetry property of the Riemann tensor. Rather than show this here, we will use the same approach to obtain a very useful result regarding the second covariant derivative of a Killing field, $M=\mathcal{K}$. Hence we write
\begin{eqnarray}
  \begin{aligned}
    &(a\wedge b\wedge c)\cdot(\mathcal{D}\wedge(\mathcal{D}\wedge \mathcal{K})\\
      &=a\cdot(b\cdot((c\cdot\overrightarrow{\mathcal{D}})\mathcal{D})\mathcal{K})-b\cdot(a\cdot((c\cdot\overrightarrow{\mathcal{D}})\mathcal{D})\mathcal{K})\\
	&-a\cdot(c\cdot((b\cdot\overrightarrow{\mathcal{D}})\mathcal{D})\mathcal{K})+b\cdot(c\cdot((a\cdot\overrightarrow{\mathcal{D}})\mathcal{D})\mathcal{K})\\
	&+c\cdot(a\cdot((b\cdot\overrightarrow{\mathcal{D}})\mathcal{D})\mathcal{K})-c\cdot(b\cdot((a\cdot\overrightarrow{\mathcal{D}})\mathcal{D})\mathcal{K}).
  \end{aligned}
\end{eqnarray}
By applying the Killing equation to certain terms in this expansion we arrive at
\begin{eqnarray}
  \begin{aligned}
  a\cdot(b\cdot((c\cdot\overrightarrow{\mathcal{D}})\mathcal{D})\mathcal{K})=&-c\cdot( (a\wedge b)\cdot(\overrightarrow{\mathcal{D}}\wedge\mathcal{D}) \mathcal{K} )\\
  &=c\cdot(\mathcal{R}(a\wedge b)\cdot\mathcal{K}).
\end{aligned}
\end{eqnarray}
In particular, if we set $c=\partial_b$ we find simply
\begin{eqnarray}\label{ricci_killing}
  a\cdot(\overrightarrow{\mathcal{D}}\cdot\mathcal{D}\mathcal{K})=\mathcal{R}(a)\cdot\mathcal{K}.
\end{eqnarray}
\section{Derivatives and the metric determinant}\label{a4}
In gauge theory gravity we have from \onlinecite{lasenby,doran} for the derivative of the metric determinant
\begin{eqnarray}
  \partial_{\bar{\mathsf{h}}(a)}\det\underline{\mathsf{h}}^{-1}=-\det\underline{\mathsf{h}}^{-1}\underline{\mathsf{h}}^{-1}(a).
\end{eqnarray}
When derivatives are present, we can invoke an orthonormal frame to give
\begin{equation}
  \begin{aligned}
  \partial_{\partial_\mu \bar{\mathsf{h}}(c)}\partial_\nu &\det\underline{\mathsf{h}}^{-1}=- \partial_{\partial_\mu \bar{\mathsf{h}}(c)}(\partial_\nu\underline{\mathsf{h}}^{-1}(\gamma_0)\wedge\dots\wedge\underline{\mathsf{h}}^{-1}(\gamma_3)\\
  &+\underline{\mathsf{h}}^{-1}(\gamma_0)\wedge\partial_\nu\underline{\mathsf{h}}^{-1}(\gamma_1)\wedge\dots\wedge\underline{\mathsf{h}}^{-1}(\gamma_3)\\
&+\dots )I.
\end{aligned}
\end{equation}
On the other hand, the Leinbiz rule produces a very similar expansion in the original expression
\begin{eqnarray}
  \begin{aligned}
  \partial_{\bar{\mathsf{h}}(c)}&\det \underline{\mathsf{h}}^{-1}=-\dot{\partial}_{\bar{\mathsf{h}}(c)}( \dot{\underline{\mathsf{h}}}^{-1}(\gamma_0)\wedge\dots\wedge\underline{\mathsf{h}}^{-1}(\gamma_3)\\
    &+\underline{\mathsf{h}}^{-1}(\gamma_0)\wedge\dot{\underline{\mathsf{h}}}^{-1}(\gamma_1)\wedge\dots\wedge\underline{\mathsf{h}}^{-1}(\gamma_3)\\
  &+\dots)I,
\end{aligned}
\end{eqnarray}
so by comparison the two must be the same up to a factor of $\delta_\nu^\mu$, or in frame-free form
\begin{eqnarray}
  \partial_{\bar{\mathsf{h}}(c)_{,n}}(\det\underline{\mathsf{h}}^{-1})_{,b}=-(n\cdot b)\det\underline{\mathsf{h}}^{-1}\underline{\mathsf{h}}^{-1}(c).
\end{eqnarray}
\section*{References}
\bibliography{bibliography}

\end{document}